\documentclass[11pt]{article} 

\usepackage{times} 
\usepackage{graphicx} 

\def\xv{{\bf x}}

\title{Self-Organized Criticality as a Phase Transition}
\author{\centerline{Mikko Alava}\\ 
INFM-SMC, Dipartimento di Fisica, Universit\'a ''La Sapienza'',\\
P.le A. Moro 2 00185 Roma, Italy and\\ Helsinki University of Technology,
 Laboratory of Physics,
HUT-02105 Finland}
\begin{document}                     
\def\xv{{\bf x}}
\maketitle                            
The original sandpile model of Bak, Tang and Wiesenfeld from 1987 has
inspired lots of consequent work and further ideas of how to describe the
birth of scale-invariant statistics in various systems and in particular 
models.
In this article the basic ingredients of self-organized criticality (SOC) are
overviewed. In line with the orginal arguments of BTW SOC is now known
to be a property of systems where dissipation and external drive maintain a
delicate balance. The qualitative and quantitive understanding of the SOC
state and deviations from it can thus be addressed, by mapping the typical
cellular automata exhibiting SOC to theories, that in fact describe critical
extended systems as such. Currently two such approaches are known, based
on the connections of SOC to variants of absorbing state phase transitions
and the physics of elastic manifolds or interfaces in random media. These
are reviewed and discussed. Finally, some open theoretical problems and
experimental suggestions are outlined.

\section{Introduction}
Usually statistical physics is on the textbook level a subject that deals with
 ``boring'' properties of matter and the associated basic thermodynamic 
quantities. There is however a modern challenge to this, that attempts to 
deal with all kinds of phenomena
with statistical physics tools.
One of the simple reasons is the discovery of many (effective) 
power-laws or ``fractals'', in the statistics of measurable quantities. This implies that within
the appropriate window the system is scale-invariant, which in the language of
statistical mechanics brings us to the field of phase transitions. The next question
that follows is then: how come so many phenomena have parameters finely tuned
to the point where the equations and laws governing them allow for ``criticality''?
Consider the prototypical two-dimensional Ising model as a example: for sure one
needs to tune the temperature to $T_c$ to reach the same effect.

The seminal papers of Bak, Tang and Wiesenfeld (BTW) started a literal avalanche
in this respect. They coined the term of ``Self-Organized Criticality'' (SOC), with
an associated set of claims, conjectures, results, and a simple model well within
grasp of anyone fluent with computers. The main idea of BTW was that such a sandpile mode would be 
effective in explaining the presence of $1/f$ -noise in Nature, a
prototypical example of power-laws indeed. The purpose of this article is to give
an overview of where the mainstream SOC understanding has lead to, from the
starting point of the 1987-1988 BTW publications \cite{btw}.
For this goal it is instructive to start with the original BTW sandpile model, for
a brief look as to what the rules and aspects of the SOC sandpiles imply. These
are often most defined as cellular automata, on hypercubic lattices of size 
$L^d$ . The
original BTW numerical results concentrated on $d = 2$. Each site $x$ has 
$z(x, t)$ grains.
There are two crucial ingredients that define the dynamics in the BTW model.
When $z(x, t)$ exceeds a critical threshold $z_c$ (a constant, here 3), the site is active
and topples. The grains (4 in the  $2d$  BTW example) are removed from x and given
to the nearest neighbors (nn). The effect of this rule is to create a 
non-linearity:
in the absence of any activity in the nn's of x the site stays quiescent, forever if
$z_x \equiv z(x,t) < z_c$ . In other words, the BTW model has $z_c$ equal to a constant,
$z_c = 2d- 1$, and the toppling rule:
\begin{eqnarray}
z(x,t + 1) = z(x, t) - 2d\\
z(y, t + 1) = z(y, t) + 1
\end{eqnarray}
where $y$ denotes all the $2d$ nearest neighbors of the site $x$.

One may now study two fundamental limits (note that clearly other scenarios
can be envisioned). Either one prepares the system with a fixed number of grains
$n_{tot} = \sum_x z_x$ , and observes what happens, with e.g. periodic boundary conditions.
Due to the presence of the nonlinearity it is easy to see that there are two opposite
limits, an absorbing state where all $z_x < z_c$ and nothing happens, and a state of
eternal activity where for all  $t$ for some $x$ $z_x > z_c$ . This defines a ``fixed energy
ensemble'' (FES) as it has been denoted in the literature. Clearly there has to be a
(phase) transition as n tot is varied between the two extremes.

However, the particular case of ``original'' SOC of BTW is created by different
conditions. What one wants to obtain is a steady state. This is obtained by 
combining open boundary conditions that are balanced with a ``drive''. The boundaries
are chosen to be such that the grains which ''topple'' out of the system are lost,
simply. Eg. in $2d$ a site which touches the boundary loses a grain out of the four it
gives out. The SOC state is now obtained by using a modern version of Maxwell's
demon: if and only if there are no active sites ($z_x > z_c$) one grain is added to a
randomly chosen site $x$:  $z(x, t) \rightarrow z(x, t) + 1$. 
Except for the presence thereof there is no ``tuning'' whatsoever.
Clearly, the effect of an added grain is either to build up $z_x$ or to launch an
avalanche if x is ``marginally stable'' if $z_x = z_c$ . The original claim of BTW was
that the sizes of avalanches - measured as the number of times that any single site
would get active after a single grain addition - would follow a power-law, as would
the duration (multiple simultaneous topplings being allows) and the area or support
on the $L^d$ lattice,
\begin{equation}
f (s) \simeq s^{-\tau_s} f(s/s_c), 
\end{equation}
with the avalanche size exponent $\tau_s$ and the cut-off 
dimension $s_c = L^{D_s}$ . The
scaling function $f$ expresses the fact that avalanches are 
limited by the scale $s_c$ and
thus decays quickly beyond that. Thus there would be a subtle balance of 
driving and dissipation in a SOC system, with power-laws arising apparently 
without any fine-tuning and hence the birth of the acronym.

The essence of SOC as from the BTW model is thus non-linearity and a 
combination of dissipation and driving \cite{SOCreviews}. 
There is an easy set of questions that follows
immediately, in particular from the analogies with other phase transitions:
\begin{itemize}
\item
universality: what kind of exponents ($\tau_s$ , $D_s$ etc.) 
would one be able to
obtain depending on the particular details of the model?
\item
 what does SOC ``mean'' in fact?
\item
 how broad is this paradigm in terms of applicability?
\item
 what is the right ``continuum theory'' for SOC?
\end{itemize}

In the rest of this article the status of the two ideas and the follow-up 
questions will be examined, in the light of recent progress on mapping SOC sandpiles
to other systems. In this, the main emphasis is on analogies with interface 
depinning and absorbing state phase transitions. Both these allow to adapt to the
ensembles (SOC, FES) in which the simple SOC automata are run, and moreover
present the essential feature of a ``nonlinearity'', of a threshold element as we will
see below. These present two different ``generic'' types of continuum limits. For
the second one, we have as the dynamical variable the activity, 
$\rho(\xv, t)$. The interface depinning variants of SOC boil 
down in most effective form as a variable
$H(\xv, t) \equiv
\int^t\rho(\xv,t)d\tau$ and its dynamical equation 
$\partial_t H(\xv, t) \dots$. The picture of
SOC is completely equivalent as to how the right ensemble is created. This will be
highlighted with examples in the next section.

\subsection{Theoretical aspecs of sandpiles}
Analogies with other systems in statistical mechanics have been around for a long
while but have not been exhausted even by now. Tang and Bak already noted that
one can perturb - on the cellular automaton level - sandpiles by driving them at a
steady, slow rate so that overlapping avalanches are not too frequent. They also
conjectured - based on ideas from equilibrium phase transitions - several 
exponent relations, to describe the correlation length and the correlation time in the
proximity of the vanishing drive rate, and most importantly identified a (correct)
order parameter, the average activity or probability that a site is active ($z_x > z_c$ in
the BTW case). The mean-field treatment results in the exponents of the contact
process or directed percolation (DP) \cite{marro}, and hence it immediately brings to the
question whether sandpiles are just an example of an absorbing state phase 
transition like DP.

For the BTW model, again, Narayan and Middleton noted that it can be obtained 
from a discrete version of an equation describing charge-density waves
(not surprising considering the history of the model as such) \cite{naramid}. 
This equation is
known another contexts and with a different noise term as the quenched 
Edwards-Wilkinson equation (QEW). It describes the dynamics of a forced 
domain wall in a random magnet. The impurities pin the interface and a driving force counteracts
the pinning. Later, Paczuski and Bottcher pointed out that a particular ``rice pile''
SOC model \cite{oslo} could be mapped also into the same system.

The presence of these two analogies (see also \cite{paczuski-etal:1996,cule:1998}) or in more concrete terms mappings 
to (continuum) systems from SOC \cite{Alava2001,Alava2002}
 makes it possible to identify firmly many answers
to the above mentioned problems, and in general to understand the critical 
properties without any conceptual confusions. These are discussed, in combination with
the missing pieces of the puzzle, in the later paragraphs. As will become clear
the nature of SOC in such typical cellular automata is a particular ensemble
version of various non-equilibrium phase transitions. In many of the early papers
analogies were drawn with the physics of equilibrium ones. Here, however, the
situation is different and one for instance is not necessarily able to define a free energy for the system, or even a 
Hamiltonian (e.g. the perennial Kardar-Parisi-Zhang
problem serves as an example \cite{Barabasi}). It also becomes more evident what the correct
renormalization group procedure is, in contrast to various real-space,
mean-field and other attempts \cite{diaz-guilera:1992,rg,tang:1988,vespignani-zapperi:1997}. This leads to an understanding
of where the differences in the numerically found scaling behavior among the
BTW, Manna \cite{manna}, Zhang  \cite{Zhangetc} etc. models originate from.

In most cases it is also obvious that the SOC state is indeed achieved only in
a particular point of the phase space, the only reason why this is not obvious to
begin with is that it becomes apparent after such connections to other models are
made clear. Thus in general there is no ``generic scale invariance'' \cite{ggrin,grin91} unless
one goes one step further: in the driven interface paradigm the system still exhibits
critical fluctuations off the SOC state, but the character of the fluctuations changes.
The simple analogy here is the thermalization of the noise in a depinning problem, and the changed interface fluctuations. 
Here one has to be careful
about the ensemble that is perturbed off the critical point: in 
translation-invariant models with a uniform dissipation rate $\epsilon$
one can observe avalanches when $\epsilon \rightarrow 0^+$
\cite{tang:1988,vespignani-zapperi:1997}
similarly to the SOC ensemble. This is of course true only when the drive
rate $h$ is adjusted so that $h/\epsilon \rightarrow 0$ 
\cite{ggrin,hwakar,vespignani-zapperi:1997}, ie. in the limit that these two
very slow scales are still ``infinitely separate'' \cite{ggrin}. If one accepts the idea that SOC
actually arises from ``normal criticality'' by some clever mechanism 
(e.g. ``sweeping the instability'' as was proposed \cite{sornette}), then 
the question still remains how this can be made to work, and
in e.g. the interface picture this becomes immediately obvious.

The original BTW model itself still attracts interest since it does have a 
powerful Abelian symmetry, peculiar of a discrete cellular automaton. As 
shown by
Dhar, Priezzhev and co-workers in a series of papers there are connections 
to spanning trees on lattices \cite{dhar99,ktitarev}, 
thus to the critical $q = 1$ Potts model 
(and in $2d$  to conformal
invariance) \cite{potts}. 
In this respect it has however lost the aura of generality that
the original papers advocated, not forgetting the long-standing 
controversies about the various 
universality classes of SOC models - related 
to the numerically determined exponents based on Eq. (3). Recent
numerical studies have even pointed out that
the BTW model shows multiscaling: the avalanches exhibit a full spectrum
of scales, whose origins are not understood \cite{Teb99}. 
Such numerical studies have 
made it apparent that there are definately different universality classes, 
as the connections to interface depinning and APT's should make clear below.

In the following models that involve strict ``extremal dynamics'', in the sprit
of the Bak-Sneppen model \cite{baksnep}, are excluded. The idea here is that one (in the
interface context) advances the site which has the largest local force. Clearly this
kind of dynamical mechanism is just another version of the demon inside the BTW
sandpile, an even smarter one since he has now to know the state of the system in
greater detail. We shall also not discuss the so-called forest-fire models 
\cite{ds92,cds94},
that have the difference that one has three states for a site: empty, a living tree,
and a burning one. For this class one has no such conservation law as 
exemplified
by the Laplacian of the interface equation or by the energy field of the 
APT class
with conservation. It is an open question how to find similar 
connections to continuum models. We also omit directed models which as such are
often more simple to understand theoretically.

\subsection{Experimental signatures of SOC}
The obvious question to ask is, what is the quantity to measure given that SOC is
in fact a ``usual phase transition''? There are two answers: either one is concerned
with avalanche behavior, like in a typical sandpile model, or then one is interested
in some quantity that would be able e.g. to establish whether the dynamics are
indeed governed by or close to something akin a SOC critical point. Given the
fact that one needs a separation of timescales it becomes clear that 
sandpile-like
behavior is not easy to conjure in an experiment. On the other hand, it is clear that
a generic phenomenon like $1/f$-noise is precisely speaking far from the typical
SOC manifestations.

Laboratory-scale experiments for this purpose were attempted soon after the
first BTW papers inspired by the ``sandpile'' idea \cite{sand}. Unfortunately, there is
little reason as to why a real sandpile should behave like a cellular automaton. In
particular, the flow of grains once started involves inertia \cite{nagel}. The ingenuity of
experimentalists then lead (in Oslo) to the discovery that one can simply substitute
with (elongated) rice grains, and a power-law avalanche distribution follows 
\cite{rice} (see also \cite{aegarter2}).

In essence, to mimick sandpiles one thus needs diffusive motion of the 
``particles'', ie. the equation of motion has to contain precisely the same ingredients
of a threshold and overdamped motion once that is exceeded. Such conditions are
provided by domain walls (DW) in ferromagnets - as we will see one can easily
define a sandpile that describes the dynamics of a DW - and vortices in type-II
superconductors.
The former can be observed by e.g. optical means or by the noise produced
by domain wall motion, known as the Barkhausen effect \cite{bark}. It is possible to
observe avalanches, with a size distribution that extends as a power-law over more
than three decades \cite{bdm,durin1,sava,urbach,ZAP-98}. The classical Barkhausen DW 
experiment corresponds however to the slowly driven, translationally invariant case
outlined above (and not to the SOC ensemble): a constant external field ramp 
corresponds to a constant velocity for the driven DW.

The recipe for superconductors is to increase the external field which pushes
vortices or flux lines into a sample (recall of type II). The Bean state 
\cite{bean,beantang} that
ensues has a conservation law (vortices), and can be controlled with the rate of
increase of the external field. One can observe avalanches, exhibiting power-law
behavior \cite{flux,niobium,aegarter1}. Indeed there are sandpile models to describe such behavior
\cite{bassler}, and
on the other hand coarse-grained lattice models that are close to sandpiles in spirit \cite{Nicodemi_2001}
(to say nothing about more complicated simulations 
\cite{nori}). In both these cases the
actual role of SOC in the phenomena at hand seems slightly secondary, however.

Avalanches have also been observed in a more promising field from the point
of view of potential applications: fracture and plasticity. While there is lots 
of evidence (in terms of acoustic emission, the energy relased by microscopic events)
for the existence of power-laws both in laboratory experiments \cite{ciliberto,oma,ae} and
in earthquakes, it is not clear at all how much these have to do with SOC. In failure
of brittle materials it is clear that the underlying processes are irreversible and have
nothing in common with the basic premises of a SOC state. In the case of 
earth-quakes the scales involved are such that it may seem more prudent to explain eg.
the classical Richter's and Omori's laws (of magnitudes and aftershock intervals)
via a suitable SOC model. However, no direct connection has been established
so far \cite{socturcotte}. Plasticity and dislocation dynamics seem better in this
respect, as a field of application of SOC ideas (see \cite{grasso,weiss} for AE evidence).
One has the possiblity of a steady-state and the system can be driven by
slowly applied external strain.

Perhaps the most enticing field where SOC concepts have been utilized is
plasma and space physics.
The crucial ingredients here are the presence of non-linearities 
(given by the character of e.g. the magnetohydrodynamic equations that
can be proposed for the Earth's magnetotail sheet), external drive 
(ionospheric activity vs. solar wind, or the solar flares vs. sun's heating), and spatial 
inhomogeneity in that the dissipation and drive do not necessarily take place uniformly. Both in
space and laboratory plasmas one can obtain measured statistics (like the particle
flux driven by the nonlinear transport at the plasma edge in a fusion device, or the
magnetospheric AE index) that can be then analyzed in terms of SOC 
characteristics. Several reviews exist that outline the subject (whether one is interested in
fusion or space aspects) \cite{charbo,ssr1,ssr2,plasma,denhel,sun1}.
There is clear evidence of power-laws in eg. the waiting
times of avalanches and energy dissipated \cite{ssr1}. 
In these contexts some of the differences to usual sandpile models are based e.g. on comparisons to (thresholded) real
signals, and on the presence of correlated and varying drive signals. Both
these make comparisons between theory and experimental or observational 
quantities somewhat hard.

\subsection{Overview}
Next we start by presenting a pedestrian picture of the connection of SOC sandpile
models to those of non-equilibrium statistical mechanics. The following section
goes on to a discussion of such mappings in more detail: the universality class of
absorbing state phase transitions with a conserved field is outlined as the first one.
Section IV considers SOC in the light of interface depinning. First the basic
background about the latter is given. Then the mapping of sandpiles to discrete 
interface equations is discussed, 
together with the implications that the various terms
and the ensemble(s) have. To complete the picture we also consider
the possiblity of extending the known SOC ensembles, by
the quenched KPZ equation.
Finally, in the conclusions we list a number of open topics for further research,
and summarize the state of the art.

\section{Two models and basic ideas}
To illuminate the differences that arise from sandpile rules and the connections to
non-equilibrium phase transitions we next discuss some examples. Consider for
this sake two models, both defined in one dimension, on a lattice with $x = 1 \dots L$.
In the first one, the rules are similar to the BTW model. Each site has $z_x$
grains. For simplicity, we start with the FES ensemble and prepare the system with
one grain per site. The difference to the BTW one is that now the thresholds $z_c (x)$
are chosen to be random. We assign to each $x$ a $z_c$ value of 2 with a probablity $1- p$
and 0 with probability $p$ after each toppling. In this way, $z_c (x)$ is an iid random
variable, if one considers its values at two $x$, $x_0$ or at the same site, separated by
toppling(s) at $x$. Then the toppling rule is for $z_c < z_x$
\begin{eqnarray}
z(x, t + 1) = z(x, t) - 2d \\
z(y, t + 1) = z(y, t) + 1 
\end{eqnarray}
where $y$ are the nn-sites of $x$.
The dynamics of the model is simple: one compares the number of grains at
a site to the local threshold. In the example above, the $z_x$ follows simply from
differences in the flux from neighbors (one grain per nn-toppling) and the flux out
due to local topplings, One can define a 'local force' as
\begin{equation}
f(x, t) = n_{in} - n_{out} - z_c (x)
\end{equation}
in terms of $n_{in}$ (grains added to site $x$ up to time $t$) and $
n_{out}$ (grains removed from $x$). The $n$-variables can be directly 
be interpreted in terms of an interface variable,
$H(x, t)$, or history one which follows the memory of all the activity at x.

In particular, the above sandpile turns out to be  the Leschhorn-Tang (LT)
cellular automaton \cite{leschhorn:1993}, 
used to simulate interface depinning for the Quenched
Edwards-Wilkinson (QEW, or Linear Interface Model, LIM) equation
\cite{qew2,nattermann-etal:1992}. The LT
automaton follows an interface $H(x)$ at each discrete time step $t_i$ 
with the equation:
\begin{eqnarray}
H(x, t_{i+1}) =H(x, t_i ) + 1, f(x, t_i ) > 0\\
= H(x, t_ i), f(x, t_i ) \leq 0
\label{lhorn}
\end{eqnarray}
where the force $f$ is in the QEW/LIM language a combination of elasticity and
a random quenched pinning force
\begin{equation}
f(\xv, t_i ) = \nabla^2 H(\xv, t_i ) + \eta(\xv, H).
\end{equation}
$\nabla^2 H(\xv) =$ (1d) $H_{i+1} - 2H_i + H_{i-1}$
 is the discrete Laplacian. The noise in the
original LIM cellular automaton is
\begin{equation}
\eta(\xv,H) = \left\{ 
\begin{array}{ll}
+1, &p \\
-1, &1-p
\end{array} \right.
\end{equation}
This choice implies that there is a average driving 
force $F = \langle f \rangle = 2p-1$, which
is the control parameter (the brackets denote ensemble averaging). 
The critical point is estimated to be at $p_c \sim 0.800$ 
\cite{leschhorn:1993}. 
Note in particular that (obvious) fact
that the FES critical point is that of the QEW universality class, 
to which we return later.

So we have discovered that we can easily associate a sandpile model with the
QEW cellular automaton. If one would use open boundary conditions this would
mean that in the SOC ensemble, with no active sites, one grain is added to a 
randomly chosen site, $z(x, t) \rightarrow z(x, t) + 1$. 
One can now look at $f(x, t)$ in this case,
and notice that there is a ``columnar'' force term $F (x, t)$ 
which counts the number of grains added to site $x$ by the external drive. 
In the original example $F\equiv 1$. In
this ensemble the non-linearity is provided by the random thresholds 
inside $f$, and
the slow drive is the same as in the QEW case, basically.

\begin{figure}
\centerline{\includegraphics[width=6cm]{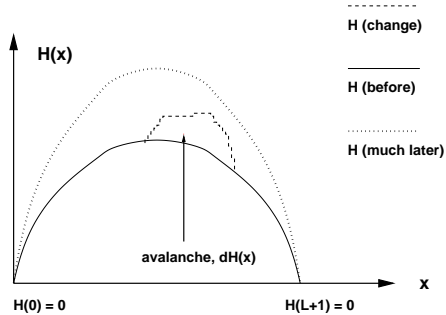}}
\caption{One-dimensional schematic example of the interface or history 
representation of a SOC model. The mean-field interface is
parabolic, which implies that in the SOC steady-state $v(x)$
is parabolic, too.
Notice the boundary conditions $H = 0$ that ensure the loss of particles 
(equalling the increased elastic energy).}
\end{figure}

The SOC ensemble works (Figure (1)) so that the local force is increased 
in steps, by adding a grain at $x$ and making $F(x, t) \rightarrow 
F (x, t) + 1$. Sometimes,
$f > 0$ and an avalanche starts. The interface moves at $x$ by one 
step, $ \delta H(x) = 1$.
In the subsequent dynamics during the avalanche the columnar force term 
$F (x)$
does not change. The right choice of the interface boundary condition is 
$H = 0$ which is to be imposed at two ``extra sites'' 
($x = 0$, $x = L + 1$ for a system of size
$L$ in $1d$). The $\langle F(t)\rangle$
increases slowly with time, so that the shape of the interface
is on average a parabola. There are now three easy-to-define questions 
following from the SOC setup: i) can one understand the ``avalanches'' 
(bursts of activity)
from the QEW depinning, with its presumably well-known exponents? 
ii) what is
the effect of the parabolic interface shape on the physics of depinning, 
if any? (this
relates to i) if things get complicated) and iii) how does the dynamics 
of such a
driven parabola compare again to that of the ``usual'' QEW, 
if it is perturbed off the critical point?

The other model is simply constructed by considering activated random 
walkers. One puts $z_x$ of these per site, and the rules are such that if 
$z_x = 1$ the walker is
frozen, and for $z_x > 1$ per unit time two walkers take off from $x$. 
This is in actual
sandpile terms the Manna model \cite{manna}, if the walkers are called ``grains''. 
We have that
$z_c = 1$, and the toppling rule
\begin{eqnarray}
z(x, t + 1) & =& z(x, t)- 2, \\
z(y', t + 1) &=& z(y' , t) + 1,\\
z(y'', t + 1) &=& z(y'' , t) + 1, 
\end{eqnarray}
with $y'$ and $y''$ two randomly chosen neighbor sites of $x$.

In the FES case (consider periodic boundary conditions) there is naturally a
phase transition located at $n_{tot} \le L$. 
Like in the QEW, the bare value of the control
parameter (here $n_{tot}/L$) at the critical point is hard to compute. 
The important
point is that the model is clearly an example of an ``absorbing
 state'' phase
transition: any product measure of states with $z_x = 1$ or 0 will do. 
Thus there
is an infinite number of absorbing states, and moreover trivially 
$n_{tot}$ is conserved.
These suffice to make it possible - and clear - that there can be 
differences to e.g.
the contact process (this would amount in the random walker language to 
a model where walkers can die out, and have off-spring, moreover 
$z_c = 0$), in the critical behavior.

In this model there is as well a memory effect: any configuration of 
grains in a local neighborhood is frozen till an avalanche sweeps over it
readjusting the $z_x$'s.
Note that the effect of this is diffusive since if a grain moves around, 
it does so by a simple diffusion process. The SOC ensemble is again 
obtained similarly to the LT 
automaton by allowing walkers to escape at the boundaries, and adding 
a walker at
a time randomly ($z_x \rightarrow z_x + 1$) if no active sites exist. 
In this case, it is easy (and
easier than in the QEW case) see why there might be subtle differences 
between the two ensembles. The average $\langle z_x \rangle$
 in the quiescent state between avalanches
is created by the average flux of grains out of the system. In 1D, 
clearly any site
for which $x \neq L/2$ there has to be a net flux of grains towards the 
absorbing boundaries.
Thus the profile of $\langle z_x \rangle$ will for sure have a gradient 
as a function of $x$, giving rise to ``finite size''
effects that are difficult to analyze.

\section{Sandpiles and absorbing states}

The classical model for a system with an absorbing state is the ``contact 
process'',
where particles diffuse, die and are born as off-spring from neighboring 
ones on a lattice. In terms of a coarse-grained density-field $\rho$
 one can illustrate its behavior with the mean-field equation
\begin{equation}
\frac{d\rho}{dt} = (\lambda - 1) \rho -\lambda \rho^2
\end{equation}
that includes the competition between the two mechanisms affecting 
the density $\rho$ \cite{marro,Haye}.
The MF-variant clearly has a phase transition at 
a $\lambda_c = 1$, with $\rho = 1-1/\lambda$ in
the stationary state. The physics version on a lattice is directed 
percolation, which has as well an upper critical dimension $d_u = 4$. 
The DP can be analyzed field-theoretically with standard coarse-graining 
techniques, and there is an extensive
body of work on the critical exponents and scaling functions etc. for 
$d < d_u$.

The exponents that are of interest here are such that they describe the 
behaviorat the critical point and in its proximity. 
The former gives rise to $\nu_\perp$, $\nu_\parallel$,  $z$, and
$\beta$. These describe in turn the temporal and spatial correlation 
scales, the dynamics
of the correlations, and the order parameter for $\lambda>\lambda_c$. 
One has the scaling relation
\begin{equation}
\bar{\rho}(\Delta, L) = L^{-\beta/\nu_\perp} R(L^{1/\nu_\perp}),
\label{dpsca}
\end{equation}
with $\Delta = \lambda - \lambda_c$ the distance to the critical point. 
$R$ is a scaling function with
$R(x) \sim x^\beta$ for large $x$. 
For $L \gg \xi \sim \Delta^{-\nu_\perp}$ we expect $\bar{\rho}\sim
\Delta^\beta$ using $\xi$ for the
correlation length. When $\Delta= 0$ we have that 
$\bar{\rho}(0, L) \sim L^{-\beta/\nu_\perp}$. Above the
critical point the order parameter has a stationary value.

Such exponents and the scaling contained in Eq. (\ref{dpsca})
would then be the goal for
an absorbing-state phase transition description of sandpiles. 
The example of activated random walkers aka the Manna model of the 
previous section makes it rather
obvious that the FES ensemble - with the mild caveat that the transition 
should be 
continuous - might be described analogously to the DP. The 
Manna universality class is not expected to be in the DP universality 
class, due to the presence of the
conservation law and the presence of an infinite number of absorbing 
states. The
questions that remain are then highly non-trivial: first, what is the right effective
field theory? Second, are there complications related to the specific character of
the SOC ensemble?

The former of these two questions can be answered exactly for a class of
stochastic models, with an infinite number of absorbing states, and with a static
conserved field (NDCF, non-diffusive conserved field class). 
It is the coupling of this field to the order parameter evolution that
creates a unique universality class \cite{Romu1,Romu2,AIPa}. 
This is a conjecture supported by extensive
simulations of a number different models \cite{Romu1,Romu2}: 
a conserved threshold transfer
process \cite{lubeck1}, a conserved lattice gas with repulsion of nearest 
neighbor particles, and a reaction-diffusion model (CRD) with two 
species of particles, $A$ and $B$. 
The exponents (in  $2d$ ) are illustrated in Table I, together with the CP
ones.
\begin{table}
\begin{tabular}{llllll}             %
\multicolumn{6}{c}{Steady state exponents $d = 2$}\\             
& $\beta$ & $\nu_\perp$ & $\beta/\nu_\perp$ & $z$ & $\theta$\\
CTTP & 0.64(1) &0.82(3) &0.78(3) &1.55(5) &0.43(1)\\
CRD &0.65(1) &0.83(3) &0.78(2) &1.55(5) &0.49(1)\\
Manna &0.64(1) &0.82(3) &0.78(2) &1.57(4) &0.42(1)\\
DP &0.583(4)& 0.733(4)& 0.80(1)& 1.766(2)& 0.451(1)\\
\end{tabular}
\caption{Critical exponents for steady state simulations in $d = 2$.
Models: CTTP: Conserved threshold transfer process, CRD: Conserved 
reaction-diffusion model, Manna:  Manna sandpile, DP: Directed percolation.
From \cite{AIPa}.}
\end{table}

The conserved reaction-diffusion model studied in \cite{Romu2}
 has the pleasant 
feature that it can be mapped exactly, using a Fock-space representation and 
creation-annihilation operators \cite{Fock,WOH}, 
into an effective action, or equivalently into a
set of Langevin equations \cite{Romu2}. 
The resulting theory has, forgetting some 
naively
irrelevant terms upon power-counting|,
a structure proposed earlier 
on phenomenological grounds \cite{dickman-etal:1998}. 
This set of equations reads
\begin{eqnarray}
\frac{\partial\rho(\xv,t)}{\partial t}
& = & 
 a\rho(\xv) - b \rho(\xv)^2 + \nabla^2 \rho(\xv,t) - \mu \psi(\xv,t)
\rho(\xv,t) \label{apt}\\
&+& 
\sigma \sqrt{\rho(xv,t))} \eta(\xv,t) \nonumber\\
\frac{\partial\psi(\xv,t)}{\partial t}
&= &D \nabla^2 \rho(\xv,t)
\end{eqnarray}
For $\rho$, this looks like a Reggeon field theory \cite{RFT}
(used to describe DP) 
which is coupled
to an extra conserved non-diffusive field, $\psi$. 
One thing is immediately clear, that
is the theory contains a term that accounts for the memory of the dynamics of
the grains in the model ($\phi$, again) 
which then couples to the actual activity $\rho$.
These fields are also often called ``energy'' (grains) and ``activity'' 
(active sites), respectively.
Due to the presence of this coupling the theory is non-Markovian.

Above, $\eta$ is a Gaussian white noise with a trivial correlator 
$\langle \eta(\xv, t) \eta(\xv', t') \rangle
\sim \delta (\xv - \xv') \delta (t-t')$. 
The parameters ($a$ and so forth) are fixed. Notice the presence
of the linear term, that orignates from the initial configuration. 
The effective noise
strength is linear in the local density, since in the coarse-graining 
the activation and
passivation of particles on the microscopic level becomes a 
Poissonian variable,
with the variance equalling the expectation. 
In particular it vanishes for $\rho= 0$.

The second equation describes how the conserved density diffuses 
upon the influence of actual activity. In the absence of such the 
dynamics is frozen. It can
be integrated out formally, to obtain a single equation for the 
activity $\rho$. More concretely
\begin{equation}
\psi(x, t) =  \psi(x, 0) +D \int_0^t dt' \nabla^2 \rho(\xv, t').
\label{intout}
\end{equation}
The first contribution in Eq. (\ref{intout}) follows 
from the initial condition, and is a
``quenched'' term: it is frozen and not affected by the presence of 
the thermal noise.
The second is a non-Markovian term. The theory as it stands has one 
easy consequence: $d_u = 4$ 
(though there is a claim by Wijland that in fact $d_u = 6$ \cite{ucd6}) in
agreement with numerics on this universality class. The problem 
is that Equation
(\ref{apt}) is difficult to renormalize due to the presence of the extra terms (in addition to
the DP-RFT) \cite{AIPa}. 
Thus one has no useful predictions as such. In the next section
we discuss the relation of SOC to depinning, in which case the corresponding 
situation in the QEW becomes pertinent. For quenched depinning the lesson is that
the quenched noise field renormalizes in a highly non-trivial fashion (in particular
from the viewpoint of the technicalities of the computations). Here an analogy
would be the correlated activity that reflects the landscape (grains or particles), at
various instances of time.

It should also be noted that the theory of Eq. (\ref{apt}) 
fails in the presence of a
number of effects - or does not hold of course for all possible sandpile models. The
BTW model has no randomness in the toppling rules, its conservation laws
play a role and the situation is different \cite{dhar99}. 
For the LIM automaton of the previous
section, the additional quenched noise clearly plays a role in the way the activity
stops (or is ``pinned''). The question what the NDCF class means
as a depinning problem will be discussed in the next section,
but it is worth underlining that the connection between the
two kinds of transitions would merit further study \cite{AM2002}. 

The SOC state in the FES case arises on a simplistic level by combining an
absorbing state with another demon (a close cousin of the BTW one), with the task
of driving the system if $\rho= 0$, everywhere. This is combined by a dissipation (like
the boundary losses) that ensures that for 
$\rho> 0$ $d\rho/dt < 0$. This is of course
just a complicated way of stating the simple rules of CA's 
(since we know that
the state is such that $\langle \rho \rangle = 0^+$ 
due to the separation of timescales with respect
to dissipation and driving). Technically this implies that 
$\langle z_x \rangle$ in the SOC case is
limited above by the critical value of the FES critical point.
This is trivially so
since the opposite would imply that $\langle \rho_x \rangle > 0$, 
locally for some $x$. The actual
scaling function of $\langle z_x \rangle ( \xv, L)$
is however not known. This is in fact one of the
fundamental questions of SOC sandpiles: if 
one can separate ``bulk'' behavior from
boundary effects.

Recent work by Dickman has shown that precisely at the SOC critical point
the sandpile avalanche properties obtain logarithmic corrections to their scaling
functions with $L$. 
This surprising finding implies that the finite size scaling in
the SOC ensemble is not easy to understand based on the translationally invariant
FES (or depinning case). This would also be expected to hold in general for the
NDCF-class (for all its models in the SOC ensemble).

Notice however again that the description in terms of the coarse-grained 
theory invites for two other scenarios: clearly the critical point can be approached by
any suitable dissipation mechanism, not only the boundary loss -one. 
A possibility is bulk dissipation (e.g. a constant probability per step that a diffusing grain
is removed). Such perturbations are closer to the FES case, since translational
invariance is restored.

\section{SOC models and depinning}
\subsection{Interfaces in random media}
Rough interfaces 
\cite{HHZ,Barabasi,Mea98,sorn}
are objects with ``self-affine'' geometry. They can
move due to an applied force, 
such as a magnetic field or a pressure difference. The essential point is
that the thermal fluctuations may be neglected, if the noise
is due to the presence of impurities or defects in the material.
Such disorder is quenched, static in time. The interface
can get locally pinned, if a local ``pin'' is strong
enough to overcome (generalize) surface tension. A characteristic
example is the quenched Edwards-Wilkinson (QEW), or linear
interface model (LIM)
\cite{qew2,nattermann-etal:1992,leschhorn:1993}:
\begin{equation}
  \frac{\partial H}{\partial t} =  \nabla^2 H + \eta(x,H) + F.
                                        \label{QEW}
\end{equation}
Here $H$ is driven by $F$, and the combination of the surface
tension and the randomness $\eta$ gives rise to the critical
behavior. In particular, depinning takes place at a 
force $F$ close to a critical value $F_c$. 
Depinned interfaces move with a velocity (order parameter)
$v\sim f^\theta$, with $f=F-F_c \geq 0$. Pinned interfaces
are blocked by ``pinning paths/manifolds'' which arise from the
quenched disorder in the environment.
\begin{figure}
\centerline{\includegraphics[width=8cm]{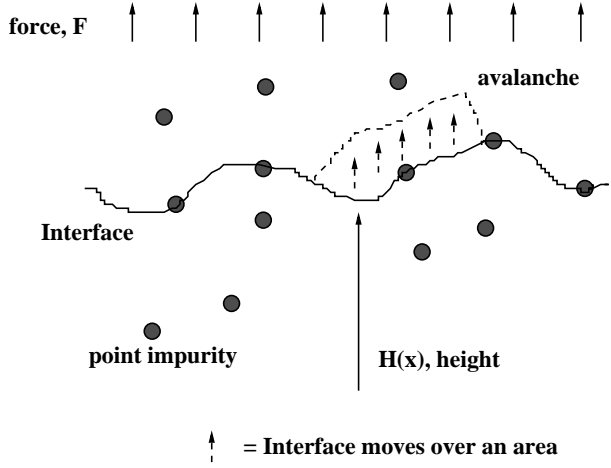}}
\caption{An interface in a random enviroment: a part
moves under the influence of $F$, the rest is pinned.
}
\label{fig:explainif}
\end{figure}

The description of kinetic roughening involves analogously
to the absorbing state phase transitions a correlation
length $\xi$.  Statistical scale invariance assumes 
(statistical) translational invariance, developing both in
time and space \cite{HHZ,Barabasi}.
The typical quantity to measure is the roughness (mean square
fluctuation) $w_q(L,t)$, where $L$ is the measurement scale.
One often observes power law scaling for its moments,
\begin{equation}
\label{w_q_def}
w_q(L,t) \equiv \langle \langle \delta H(\xv)^q \rangle
\rangle^{1/q} \sim L^{\chi_q} \mbox{ for
  } L < \xi(t)
\end{equation}
saturating to a constant for $L$ larger than $\xi$ (the inner brackets imply
averaging over $\xv$, and the outer an ensemble average). $\chi_q$ is
called $q$-th order roughness exponent, allowing for {\em multiscaling}
should the $\chi_q$ differ from each other (as for ``turbulent
interfaces''\cite{Krug_1994}, when the correlations
are dominated by the  largest height difference between two neighboring points,
$\Delta H = |H_{x+1} - H_x|$
 \cite{Krug_1994,Asikainen_2002,Mitchell_2002}).

Usual self-affine scaling means  $\chi_q \equiv \chi = \mbox{const}$,
and  $\xi(t)$ increases as $\xi(t) \sim t^{1/z}$,
with $z$ the dynamical exponent. The maximal extent of interface
fluctuations 
\begin{equation}
\label{wt_scaling}
w(t) \equiv \lim_{L \to \infty} w(L,t) \sim \xi(t)^\chi \sim
t^{\chi/z} \equiv t^\beta,
\end{equation}
which defines the exponent $\beta$, combining into
a scaling form
\begin{equation}
\label{scaleform}
w(L,t) = \alpha(t) \; \xi(t)^\chi \; \mathcal{W} \left( \frac{L}{\xi (t)}
\right).
\end{equation}
with the scaling function $\sim x^\chi$ for $x < 1$ and a constant for 
$x \gg 1$. If the amplitude $\alpha(t)$ increases, so-called anomalous scaling
ensues \cite{Lopez_1997}, and if $\chi > 1$ ''superroughness''
\cite{Lopez_1997,Lopez_1999}.
The LIM in $1d$ is an example
\cite{leschhorn:1993,Leschhorn_1993b,Leschhorn_1994,Leschhorn_1996},
and below we will discover that several SOC automata in the interface
description have this property as well.
$\chi$ can also be described by the structure factor,
or spatial power spectrum
\begin{equation}
\label{structfact}
S({\bf k},t) \equiv \langle | h({\bf k},t) |^2 \rangle \sim
\Biggl\{
\begin{array}{ll}
\! k^{-(2 \chi + d)} & \mbox{for }k \gg 1/\xi(t) \\
\! \xi(t)^{2 \chi + d} & \mbox{for }k \le 1/\xi(t)
\end{array}
\end{equation}
where $h({\bf k},t)$ denotes the spatial ($d$-dimensional) Fourier transform
of $h({\bf x},t)$. Also, the height difference correlation function
$G_2({\bf x},t) \equiv \langle | h({\bf x},t) - h(0,t)|^2 \rangle^{1/2} =
\int_{\bf k} S({\bf k},t) \left( 1 - \cos({\bf k \cdot x}) \right)$
can be used, but note that it reflects only the local
roughness exponent $\chi_{\rm loc}$ and thus for
superrough interface, with $\chi > 1$, one needs $S$ (for $L$ fixed).
For temporal quantities the analogue is naturally the power spectrum,
in the stationary state,
$S({\bf x},\omega) \equiv \langle | h({\bf x},\omega) |^2 \rangle$
(which has obvious problems in the SOC state, due to the separation
of timescales), or 
correlation function of the $q$th moment of height ``jumps'' over a temporal
distance $t$
\begin{equation}
\label{Ct}
C_q({\bf x},t) \equiv \langle | h({\bf x},t+s) - h({\bf x},s)|^q \rangle^{1/q}.
\end{equation}
Generally one finds an increase $C_q({\bf x},t) \sim t^{\beta_q}$ 
over ``short''
time distances $t$. In the case of (conventional) scaling it is related
to the early time increase of the width, $q$-moments $\beta_q \equiv \beta =
\chi/z$. It is easier to use than $w(t)$ at early times.

The main universality class of the QEW, Eq. (\ref{QEW}), is reached
for random bond and random field disorder (these flow in the RG into
the same fixed point). The RF noise correlator reads
\begin{equation}
\langle \eta(x,h(x,t)) \eta(x',h(x',t')) \rangle = \Delta
(h(x,t)-h(x',t')) \delta(x-x')
\end{equation}
where $\Delta(u)$ is in practice taken as a delta function,
$\Delta (h(x,t)-h(x',t')) \rightarrow \delta (h(x,t)-h(x',t'))$.
If the velocity $v$ is finite, the average motion and
the fluctuations separates, hence for
$l_{th} > (vt)^{1/\chi}$, 
\begin{equation}
\delta (h(x,t)-h(x',t')) \rightarrow \delta(v(t-t')+\delta h) 
\rightarrow \frac{1}{v} \delta(t-t')
\label{crossover}
\end{equation}
the noise becomes thermal, with the strength
$\tilde{\Delta} = \Delta_0 / v$, and
$\langle \eta(x,t)) \; \eta(x',t') \rangle = 
2 T \delta(x \! - \! x') \delta(t\! - \!t')$.
The QEW becomes then the normal Edwards and Wilkinson one,
for a surface relaxing by surface tension \cite{Barabasi}.

The QEW develops critical correlations in the vicinity of the critical point,
$F_c$. The standard analysis of the problem is the functional renormalization 
group method. One-loop expressions for the exponents are found
in papers by Nattermann et al. and Narayan and Fisher 
\cite{nattermann-etal:1992,qew2}.
The analysis has been pushed recently further by Le Doussal, Wiese
and collaborators \cite{Doussal} illuminating several 
open issues.
The main fixed point is the RF one, but for {\em correlated} noise
there is the possibility of continuously varying exponents.
The upper critical dimension is $d_c=4$, above which
mean-field theory applies. 
Several exponent relations exist close to the critical point,
like $\theta=\nu(z-\chi)$ for the velocity
exponent, and  $\chi + 1/\nu = 2$. This,
together with the $\theta$-exponent relation, tells 
that there is only one temporal and one spatial
scale at the critical point. 

For the RF fixed point, the 1-loop functional RG results
are $\chi = (4-d)/3$, and $z = 2 - (4-d)/9$.
Later work by Chauve, Le Doussal,
and Wiese \cite{Doussal}
yields (with $\epsilon \equiv 4-d$)
$\zeta =\frac{\epsilon}{3} (1 +0.14331 \epsilon +\dots)$.
In particular in $d=1$, this means that the RG beyond
one loop is better able to adhere to the numerical
LIM results 
\cite{leschhorn:1993,nattermann-etal:1992,Rosso_2001}
with a superrough interface ($\chi_{{\rm QEW},d=1} \sim 1.2 \dots 1.25$).
Note that the depinning problem can also be discussed
in the {\em constant velocity} ensemble, which allows
for the presence of avalanches (since parts of the interface
are pinned close to $F_c$) in the presence of translational
invariance \cite{Nara,Lacombe_2001}. 

There is one ``easy'' case in which one can understand
SOC with depinning, ``at once''. This comes in the form
of the $1d$ Oslo ricepile model \cite{christensen-etal:1996},
which operates on the idea of having slopes $z_i = n_i - n_{i+1}$,
at each site, which are compared with a critical slope $z_c$, either
one or two. The dynamics is such that a overcritical site
gives one grain to its right neighbor, and thus
decreases $z_i$ by $2$, while its nearest neighbors' increase one.
After each toppling $z_c$ is redrawn from a probability distribution.
If there are no active sites, the avalanche stops and the model
is driven at the boundary ($i=1$).
Paczuski and Boettcher noticed that this can be mapped,
(since $z_i$ changes as an elastic force term) into a
boundary-driven LIM \cite{oslo},
and conjectured that the discrete equation is in the
LIM class.
Recently Pruessner has shown that one can do the mapping
in a slightly different way, which allows to get around
the delta-functions involved in this particular model
systematically \cite{pruessner}.

\subsection{Mapping SOC to interfaces: the noise}
One would like to describe sandpiles with
the dynamics of the ``interface'', of the history of
the sandpile, as in the discrete version of a QEW of
Section II, the Leschhorn automaton. For this purpose
it is also instructive consider some other models than
the ``LIM'' sandpile, the BTW and the Manna ones \cite{Alava2001}.

The Zhang model \cite{Zhangetc} resembles both the BTW model and the Manna
model in that the topping rule is a combination
of deterministic and stochastic factors. It is defined in
with a continuum 'energy'
$z$, with the dynamics
\begin{eqnarray}
        z(x,t+1) & =  & 0, \\
        z(y,t+1) &  = & z(y,t) + z(x,t)/2d, 
\end{eqnarray}
and $z_c = 2d-1$. These imply that the 'energy' of a critical site
is divided equally between the neighbors.  
The drive in the Zhang model is usually, in the SOC case, 
implemented so that the energy is increased
by small, finite amounts at a random site $x$. Or, one picks
the site closest to $z_c$, since it will
typically sooner or later be the one to reach criticality first. 
In the Zhang model the detailed history of topplings
is of importance making serial and parallel dynamics for active
sites to differ. Note that the Leschhorn (QEW) automaton and the BTW 
one are Abelian (as is the Manna model, in the general sense), so that
it explicitely does not matter if parallel or serial topplings
are used.

Another widely studied model is the Olami-Feder-Cristiansen (OFC) model
\cite{OFC},
usually studied in the following deterministic version.
The system is prepared with an initial (continuum) energy profile $z(x,0)$
and a threshold $z_c$ is chosen. Now, define a distribution parameter
$\alpha$. With the aid of $\alpha$ the OFC model reads if $z_x > z_c$
\begin{eqnarray}
        z(x,t+1) & =  & z(x,t) - z_c, \\
        z(y,t+1) &  = & z(y,t) + \alpha z_c,
\end{eqnarray}
with $\alpha \leq 1/4$ in  $2d$ . The point of dissipation
($\alpha<1/4$) is that one can achieve a steady-state even
if the open boundary conditions are replaced with e.g. periodic ones.
The same trick can also be applied to any of the previous models, where
one can do away with dissipative boundaries, and substitute for that
with a small but finite removal rate of grains that topple
\cite{vespignani-zapperi:1997}.

A third case, very similar to the Manna, is 
a microscopic model in the NDCF class \cite{Romu2}.
Consider a two-species reaction-diffusion 
process, with particles of types  $A$ and $B$ involved. 
At each site $i$, and at each (discrete) time step
the following reactions take place:
\begin{eqnarray}
B_i       & \rightarrow & B_{nn}       \, ,\,\, r_d (\equiv 1)\\
A_i + B_i & \rightarrow & 2 \times B_i \, ,\,\, r_1\\
B_i       & \rightarrow & A_i          \, ,\,\, r_2 .
\end{eqnarray}
The $A_i$, $B_i$ denote particles of each kind at site $i$.
$r$'s are the probabilities for the microscopic processes
to occur: diffusion, $r_d$, activation $r_1$, and passivation $r_2$.
Fix $r_d=1$, implying that, after having the chance to react,
$B$ particles diffuse with probability one, and a phase 
boundary follows between the active and absorbing phases in terms of the 
$r_1$, $r_2$ probabilities.
For occupation numbers $n_{A,i}$, $n_{B,i}$ per site,
since the $A$'s are non-diffusive, there is
an infinite amount of absorbing states defined by $n_{B,i} =0$ for
all $i$, with $n_{A,i}$ arbitrary, similarly to the Manna model \cite{AM2002}.

The idea of how to map the dynamics of sand in these
models into a noise term is similar to
the microscopic arrival-time mapping which
connects directed polymers in a random medium to the temporal 
behavior of a roughening interface, governed by
the Kardar-Parisi-Zhang equation \cite{Barabasi}. 
It is also a cousin
of 'run-time statistics' by Marsili et al., 
which maps the quenched disorder in say invasion percolation 
to an effective memory term at each location \cite{marsili}. 
The ensuing noise in a sandpile LIM arises 
from annealed to quenched disorder mappings.

Essentially, one constructs the ``local force'' ($f$)
by looking at how $z_x$ deviates from its {\em expected
value} at the time of toppling. This is obtained by considering the net effect
of the sandpile rules, by decomposing via a projection
trick the expected 
incoming grain or energy flux as
\begin{equation}
n_x^{in} = \bar{n}_x^{in} + \delta n_x^{in},
\end{equation} 
where $\bar{n}_x^{in}$ measures the expected flux into site $x$ up to
time $t$ and $\delta n_x^{in}$ is the deviation.
The average part can be used to construct
a Laplacian, while $\delta n_x^{in}$ generates a 
noise term
\begin{eqnarray}
   \tau(x,H) &=& \delta n_x^{in}  \equiv  n_x^{in} - \bar{n}_x^{in} \nonumber\\
        &=& n_x^{in}- \nu \sum_{x_{nn}} H(x_{nn},t).
\label{deltas}
\end{eqnarray}
The sum over the nn's in Eq.~(\ref{deltas}) is the average
grain flux into $x$ due to each toppling of a nn, computed
exactly at the time of toppling at $x$ at height $H$. 
For the reaction-diffusion model one needs to 
define $H$ by the integrated activity of $B$-particles,
so that one $B$ diffusing out of $x$  means a ``toppling''.
Then the flux fluctuates since $B$'s diffuse, randomly,
while the reactions between the $A$ and $B$ species are
accounted for every time $B$ particles leave the site $\xv$.
Ie., one considers $n_A$ and $n_B$ when the site becomes 
active and a particle diffuses out. Then
$n_{tot}(x,H)= n_A (\xv,H) + n_B (\xv,H)$.

\begin{figure}
\centerline{\includegraphics[width=6cm]{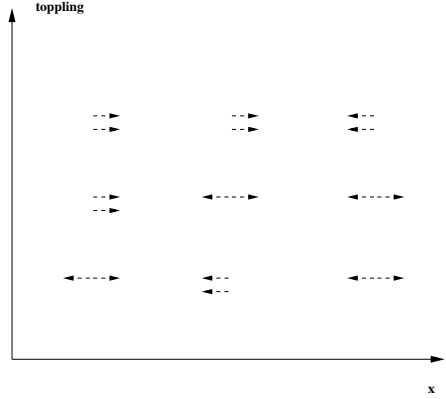}}  
\vspace*{5mm}
\caption{How the randomness maps to an interface noise
field in the Manna model in 1d: each toppling implies one
expected grain at a nn, but can give zero or two instead
(contributing to $\delta n_{nn}^{in}$).}
\label{fig:explain:M}
\end{figure}

In the case of the Manna model,
each nn-toppling contributes $1/d$ to $\bar{n}_x^{in}$ and
thus $\nu = 1/d$ while for the others the average flux is
dependent on the average energy at the time of toppling,
\begin{equation}
\delta n_x^{in} = n_x^{in} - z_{av,c} \sum_{nn}H_{nn}.
\end{equation}
Here $z_{av,c}$ is the above 
average energy in the Zhang and OFC models,
contrary to the Manna model this is not known a priori. 
Another way to write 
the fluctuation term is $\delta n_x^{in} = \sum_{H_{nn}} 
\delta z_{H_{nn}}$, ie. the fluctuations consist of
sum of the differences between the real amount of energy
of the toppling neighbors and the expected average energy.
This makes it evident that though the Zhang and OFC model will have
the same columnar noise from the drive as the others, another component
is born out of the the sandpile rules.

In the RD-model, 
the noise takes into account that since after a toppling
a passive site can get activated only after a nn topples,
and since the reactions make $n_B$ fluctuate while it is
non-zero. Thus {\em if and only if} a site does not topple
during a time step, it implies that there is an effective
threshold so that $n_B - n_c <0$ or that $n_B=0$. These
give rise to a $\tau$-distribution, that
depends upon the total number of
particles after the preceding toppling and the 
microscopic dynamical rules at a site. The
immobile grains $n_A$ imply a ``pinning force'', as
a large $n_A$ for a constant $n_{tot}$ reduces
the probability to topple.

The noise $\tau(x,H)$ is dependent on the choice of dynamics 
or the toppling order, and this becomes important if the
model is not Abelian (the Zhang model)
Keep in mind, that 
the mapping describes any particular sandpile with an interface, 
which follows exactly the same dynamics. For any change of dynamics
this implies that the noise field has to change as well. 
The $f$ of the discrete interface equation reads thus
\begin{equation}
f = \nu \,\, \nabla^2 H + F(x,t) - z_c(x,H) + \tau(x,H)
\end{equation}
allowing for varying $z_c$ for the sake of generality,
with $\nu = 1/d$ for the Manna.

\begin{figure}[tb]
\centerline{\includegraphics[width=6cm]{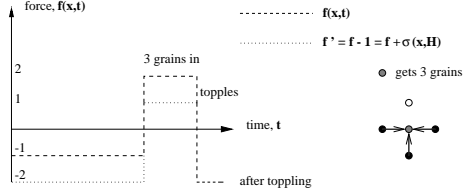}}  
\vspace*{5mm}
\caption{Rescaling of the force $f$ and an example
of how the $\sigma$-noise ensues (three grains added simultaneously)
due to ``extra'' grains present when the site topples.}
\label{fig:explain}
\end{figure}

The other source of ``SOC'' noise arises since 
the implied step-function, $\theta(f)$, in Equation~(\ref{lhorn})
forces the corresponding interface to move only forward,
so that the velocity $v \equiv \Delta H / \Delta t$ is either 0 or 1
\cite{leschhorn:1993}. This can also considered as an origin
of noise, a term $\sigma(x,H)$. An illustration is shown in
Fig.~\ref{fig:explain}.
On the avalanche timescale $f<0$ at site $x$ and 
$f$ increases until at time
$t-1$ one or more neighbors topple resulting in $f>0$.
Then site $x$ topples at time $t$. 
Thus the sandpile rules result in an {\em effective\/}
force $f'\equiv1$, acting on $H$ as
$\Delta H/\Delta t = f' \, \theta(f) \equiv f' \theta(f' )$.
The relation between $f$ and $f'$, valid for as long as
$H$ is constant, is
\begin{equation}
	f'(x,H) = f(x,H) + \sigma(x,H), \,\,\,\,\,\,\,\,\,
       \sigma(x,H) = 1 + z_c(x,H) - z(x,t^*)
				\label{f-renorm}
\end{equation}
where $t^*$ is the time at which site $x$ topples. 
This means specifically that one can construct $\sigma(x,H)$
as a quenched random variable, at $t^*$, as if it were noise
included in a QEW from the very start. 
It is computed from the difference between $f$ and $f'$
when $x$ topples. 
$f'$ and $f$ are time-dependent,
they change as grains are moved, or the Laplacian changes.
The equation $\Delta H/\Delta t = f' \theta(f' )$
can finally be viewed as the discretization of the continuum equation
$\partial H / \partial t = f'$.

\subsection{Sandpiles at criticality}

Such mappings result in
the discretized interface equation
\begin{eqnarray}
	\frac{\partial H}{\partial t}  &=& 
	\left(\nu \nabla^2 H + \eta(x,H) + F(x,t) +\sigma(x,H)\right) 
	\nonumber\\
	&& ~~~ \times \theta\left(\nu \nabla^2 H + \eta(x,H) + F(x,t) 
                                      +\sigma(x,H)\right) ,
					\label{qew_eq}
\end{eqnarray}
with the surface tension $\nu$ depending on the model,
and with $\eta(x,H) = - z_c(x,H) + \tau(x,H)$.
This is the central difference discretization of the LIM, 
and parallel dynamics are usually assumed.
The physics of such an equations contains the Laplacian character 
of grain dynamics, which corresponds to the elasticity.
topplings map exactly to an elastic force. The sandpile
dynamics (a toppling) is translated into a force
$\eta + F$, that manages to overcome a pinning force, 
so that the interface moves by one step.
The rules of the individual sandpiles are embedded
in the noise variables [$F$, $\eta(x,H)$] 
as do the details of the dynamics [$\sigma$].
The ensemble (SOC, FES or depinning etc.) is reflected
in the boundary conditions, eg. $H=0$ at the boundaries.
The character of the criticality is therefore determined
by a combination of these two: noise, and ensemble.

The theoretical understanding of the relevance of the
noise terms implies that pertubations of the type $\eta$
are {\em relevant}, and lead to the establishment of
unversality classes that depend on the RG flow of the noise
field upon rescaling. In particular, there is the LIM/QEW
class. As discussed earlier the NDCF/Manna class defines another
one, characterized by a different set of exponents.
The upper critical dimension of the LIM is $d_u = 4$ for all
cases,  however, because of the Laplacian in 
Eq.~(\ref{qew_eq}). Table 2 lists some recent
numerical results of Chate and Kockelkoren \cite{chate}. 
The BTW model presents
a different story, due to the fact that the continuum
and discrete versions differ (as we discuss below in the
context of the $\sigma$-term). Notice that its continuum
version is directly solvable, and corresponds to the
deterministic relaxation of an initial height profile
{\em in the absence of noise}, of any type. This is since
the height profile can be recast into a form that accounts
for $F(x,0)$. For the OFC and Zhang models the situation
is not so clear - they both present {\em deterministic}
dynamics during the avalanches, ie. an absence of explicit
noise terms of the type $z_c$, but $F(x,t)$ is still
present.

While $\eta(x,H)$ term in the LIM case has an 
exact equivalence $z_c(x,H) = \eta(x,H)$, implying
point disorder, the noise field for the NDCF class, including
the Manna model has a noise field $\tau(x,H)$
which is point-like but correlated. This is since
$\tau$ is {\em conserved},
\begin{equation}
\sum_{x,H(x)} \tau(x,H) \equiv 0
\end{equation}
in the case of a closed system (periodic boundary conditions,
e.g. the fixed energy ensemble), and approximately for the SOC one. 
The average of $\tau$ is zero, $\langle \tau \rangle = 0$. 
For the Manna model at any particular $x$ the increments of
$\tau$ are random walk -like, since the neigboring sites'
topplings constitute a random point process.
Define the noise two-point correlation function by 
\begin{equation}
C(\Delta x, \Delta H)=
\left<\tau(x +\Delta x,H+\delta H) \tau(x,H) \right>,
\end{equation}
where an average is performed over the 'disorder', or
in the sandpile sense many toppling histories. The
random walk-character of the $\tau$ means that 
\begin{equation}
C(\Delta x, \Delta H) \sim f(\Delta x) (\Delta H)^{1/2}
\end{equation}
since the increments are uncorrelated at a singel $x$, and
$f$ denotes a correlator in the $x$-direction. This kind
of noise correlations are similar to the ABBM-model of
Barkhausen noise. The character of the $x$-part of the
noise correlator for the Manna-model is more subtle
(note that one
might also study such a LIM, but with $\tau(x,H)$ changing after 
each toppling with uncorrelated increments).
In the SOC ensemble the interface will be parabolic, in such
a way that at a constant $H$ the various noise values 
$\tau(x,H)$ decorrelate so that one should perhaps
measure the noise correlator along the interface.
Since it is the sandpile dynamics that gives
rise to $\tau$, its local value should be related in a
non-trivial way to the local interface height. Even more
simply, if a site gets by chance very few grains
compared to the expectation, then it does not topple very
much ($H < \bar{H}$) and $\tau(x,H)$ is negative.
In the FES case $f$ can be defined and measured e.g.
numerically in the standard fashion, but $C \sim f$ should
decay quickly since neighbors have a weak influence
on each others' noise values $\tau$ (see Fig.~\ref{fig:explain:M}).
They correlate in the noise strength
since the $H$-part of the correlator is dependent on
the interface height.
The case of the RD model is slightly more complicated, though
the noise from the diffusive flux is 
the same as for the Manna, explaining qualitatively
the existence of such a NDCF universality class.
The off-critical behavior of such models
couples directly the noise field, unlike in a LIM-automaton,
and might have to do with the observation of slow relaxation
in the Manna model \cite{slowrel}.

Keep in mind that what matters in determining 
avalanche properties is the effective disorder that
influences its spreading. In the models with $\tau$-noise
this means that increments of $\tau$ stem from randomness
in the rules (and the noise correlator $C$ above measures
its feedback on the avalanche dynamics).
For other models (BTW, Zhang, OFC without additional
disorder) the avalanche dynamics does not experience
any randomness ``during'' the avalanche,
but are set by the initial properties of the system.
Eg. though for Manna $\tau$ is conserved,
in the case of the Zhang model (say) the average $\tau$ during an
avalanche can be different from zero, though its
ensemble average is zero. It is also possible to 
drive a SOC-interface with strong enough randomness
so that the interface tension vanishes. In sandpile
language this can be achieved by making e.g. the toppling
thresholds increase (even without bounds). Then a natural
possibility is a cross-over to directed percolation,
precisely since the diffusive grain dynamics is changed,
by the extra noise \cite{socdp}.

The fixed energy ensemble
\cite{tang:1988,dickman-etal:1998}
corresponds to ordinary depinning, ie. $H(x,t=0)=0$,
$F(x,t)=F(x,0)$,
and periodic BCs, with no dissipation.
One needs to tune the control parameter $\Delta F=
n_{tot}/L^d$, to obtain criticality, to $\Delta=0$.
The usual ``random preparation'' of the grain configuration
corresponds to a spatially dependent force profile $F(x,0)$,
which might give rise to some memory effects, depending on
its exact (columnar) form \cite{transient}. In the same vein,
``microcanonical'' simulations \cite{chessa-etal:1998prl} one has
dissipation operating on the slow time
scale with exactly the same rate as $F(x,t)$.
Thus microcanonical simulations correspond to
fixed energy simulations with a specific initial configuration: 
after each avalanche, the time is reset
to zero, the force is replaced with $F \to F +\nabla^2 H$, and
the forces at $x'$ ($x''$) are increased (decreased)
by one unit where $x'$ and $x''$ are randomly chosen sites.
Finally the interface is initialized, $H\equiv 0$.
Recent simulations of the NDCF/Manna (or Conserved-Directed Percolation,
C-DP) class in the depinning ensemble have revealed, that the
scaling exponents are in  $2d$  very close to LIM, while in
$1d$ and $3d$ differences ensue \cite{chate,fes2,lubeck1}.
In particular, there is the $\kappa$-exponent that relates
the local and global roughness exponents ($\chi_{global} = 2\kappa + \chi_{loc}$,,
as a clear signal of anomalous scaling (and in $1d$ of
superrough behvior) \cite{chate,fes2}; the . The BTW model is in a different class
as such, due to the extra symmetries, and as expected based
on the presence of noise in the other models \cite{btwfes}.

\begin{table}
\caption{\label{tab-c-dp} Critical exponents of the  C-DP/NDCF class
in $1d$,  $2d$, and $3d$. The corresponding values for DP and LIM are also given
for reference.
From \cite{chate}.}
\begin{tabular}{clllll}
$d$ &  &    $\delta$     & $z$           & $\beta$       & $2\kappa$ \\ \hline
  &{\sc DP} &  0.1596      & 1.58          & 0.2765        & 0.84(1)    \\
1 &{\sc C-DP} &{ 0.140(5)}& { 1.55(3)}  & { 0.29(2)} & { 0.86(1)} \\
  &{\sc LIM}  &  0.125(5)    & 1.43(1)       & 0.25(2)       & 0.35(1)\\ \hline
    &{\sc DP} &    0.451     & 1.76          &  0.584        & 0.56(2)    \\
2 &{\sc C-DP} &{ 0.51(1)}  & { 1.55(3)}  & { 0.64(2)}  & 0.50(2)    \\
  &{\sc LIM}  &  0.50(1)      & 1.55(2)        & 0.63(2)        & $0^+$ \\ \hline
    &{\sc DP} &  0.73        & 1.90          & 0.81          & 0.30(5)   \\
3 &{\sc C-DP} &{ 0.88(2)} & { 1.73(5)}   & { 0.88(2)}  & $<0.2$    \\
  &{\sc LIM}  &  0.77(2)     & 1.78(7)         & 0.85(2)       & 0    \\
\end{tabular}
\end{table}

For open boundary sandpiles, the right interface boundary
condition is $H = 0$ which is to be imposed at ``extra sites''
($x=0$, $x=L+1$ for a system of size $L$ in $1d$). In the SOC
steady-state the Laplacian increases, because of the parabolic
shape for the the toppling profile $H(x)$ (see also \cite{Barrat};
$\rho$ in the APT-picture naturally equals the same).
This is compensated by the ever-increasing $\langle F(t) \rangle$,
or the addition of grains by the SOC drive.
Writing now $\langle F \rangle \equiv ft$,
and $H \equiv H_x t$, one has the Poisson equation 
\begin{equation}
\nu \nabla^2 H_x = -f_x 
\label{mfinterface}
\end{equation}
with the constraint that $H = 0, x \in \partial x$,
and the source term $f_x = 0, \partial x$ and $f_x = f$,
elsewhere. This formulation of the 'history' of topplings
in a sandpile is just another way of looking at the 
lattice Green's function for any sandpile,
noticed by Dhar in the case of the BTW long ago \cite{Dhar}.

Concerning the fluctuations, at the depinning
critical point an interface has a diverging response.
In the sandpile language the average
avalanche size diverges with system size, 
\begin{equation}
        \left< s \right> \sim L^2
			\label{eq:<s>}
\end{equation}
which follows also from the fact that each added grain will 
perform (an effective random walk) of the order of $L^2$ 
topplings before leaving the system,
independent of dimension \cite{tang:1988,kadanoff-etal:1989}.
In Eq.~(\ref{mfinterface}) the rate of divergence is reflected
in $\nu$, in that it measures for each
particular choice of rules how the above scaling
can be interpreted as an equality, since
the total number of topplings per grain added
is dependent on $\nu$.
The linear relation between $\nu$ and the
interface velocity does in no way reflect the actual
universality classes of the models. It just manifests
the fact that the critical point of such models
exhibits ``anomalous diffusion'', as hinted by
the LIM dynamical exponent $z \neq 2$. Hence
in the interface representation such glassy
response is easy to understand \cite{singdiff}.

Recall that in the SOC steady-state the probability to have avalanches of 
lifetime $t$ may have the distribution
$
	P(t) = t^{-\tau_t} f_t(t/L^z).
$
$P(s)$ was mentioned in the Introduction, and
$s\sim t^{D_s/z}$ and $z(\tau_t-1)=D_s(\tau_s-1)$.
The linear size $\ell$ related to $s$ scales as $s\sim \ell^D$
and the (spatial) area as $\ell^{d}$ (for compact avalanches)
with $\ell$ the linear dimension.
Thus, $\tau=2-2/D$ and $\tau_t = 1 + (D-2)/z$.
From the interface depinning picture we now obtain directly
that the ``volume'' of an avalanche should scale as $d+\chi$,
and that the average area scales as 
$\left< \ell^d \right> \sim L^{1/\nu}$.
The former implies the important scaling relation 
\begin{equation}
\tau_{s} =2-2/D = 2-2/(d+\chi).
\label{taufromdepinning}
\end{equation}
This holds {\em if and only if} simple self-affine geometry
can be expected. The avalanche exponents of SOC models
are notoriously hard to measure, making it for a long
time apparent that e.g. the BTW and Manna models would
be in  $2d$  the same universality class, with $\tau_s \sim 1.27$
for instance \cite{chessa-etal:1998,biham,simulations}. 
In the light of connection to interface depinning
and (Manna, only) to NDCF class this seems fortuitious at once.
The constant-velocity ensemble is a ``good'' one for
checking out Eq.~(\ref{taufromdepinning}), and indeed
for the LIM class the expected avalanche scaling can
be found \cite{Lacombe_2001}).

For the SOC case, recent simulations have unearthed
two fundamental facts. First, the {\em dissipating}
avalanches can have interesting properties of their
own \cite{drossel}.
In the BTW model this is related to the fact
that the avalanche stops at once when the boundary is
met, while for the other models this holds as a strong
trend \cite{ktitarev}.
More importantly, the existence of {\em multifractal}
behavior has been demonstrated, also in the BTW, by Stella
and coworkers \cite{Teb99}. This presumably
follows from the strong symmetries present in it; one can also
measure many interesting quantities in the ``wave picture''
\cite{dhar99,Pri96,Pac96}.
of the BTW, which are naturally hard to relate to any
scaling theory related to interfaces or absorbing states.
 In concrete terms,
the multiscaling indicates that for e.g. the avalanche
sizes various momenta exhibit 
different scaling exponents $\tau_{s,q}$, as a function
of the pile size $L$, $\langle s^q \rangle \sim L^\tau_{s,q}$. 
For the Zhang model, Pastor-Satorras
and Vespignani have used the same technique to point out
the lack of any clear scaling whatsoever. This model can
combined with randomness, in which case the scaling attained
is that of the NDCF/Manna-class. Likewise it seems possible 
to pertub a SOC state {\em continously} (in terms of the
effective scaling exponents) between the Manna and the
Zhang model endpoints \cite{lubeckcrossover}. 
This seems to follow naturally
in the light of change in the effective noise acting on the interface
representing the sandpile. In the same vein, the existence
of strong, Manna-type noise is a relevant perturbation and
explains the cross-over to similar scaling exponents, if
such randomness is added on the top of the other sandpile
rules \cite{pastorzhang}.

\begin{figure}[th]
\centerline{\includegraphics[width=0.9\linewidth]{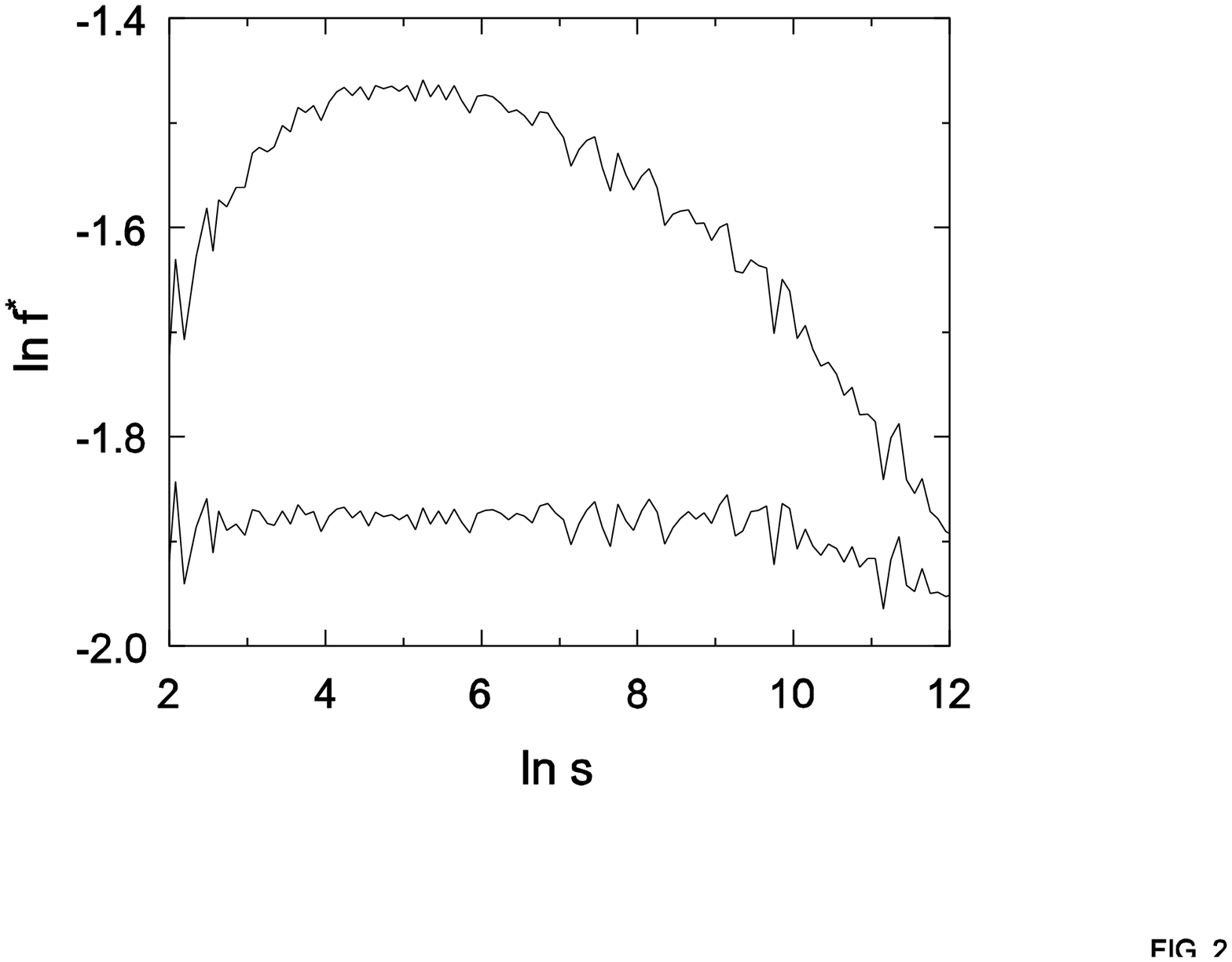}}
\caption{Plot of $f^* = s^\tau P_s(s)/(\ln s)^\gamma$
versus $\ln s$ for the data shown in Fig. 1. Lower curve:
best-fit for $3 < \ln s < 10$ using $\tau = 1.385$ and $\gamma = 0.672$;
upper curve: pure power-law fit using $\tau = 1.25$.
From \cite{dickmannotpl}.}
\label{mannanohope}
\end{figure}

Recent work by Dickman has shown however that more
fundamental surprises can be found in the NDCF avalanche
properties \cite{pastor}. Fig.~(\ref{mannanohope}) depicts
the result of careful numerical simulations,
demonstrating that one needs systematic logarithmic corrections
to describe the $P(s)$-distribution \cite{dickmannotpl}. That is,
\begin{equation}
P(s) =  s^{-\tau_s} (\ln s)^\gamma f_s(s/s_c),
\label{logcr}
\end{equation}
where $\tau = 1.386$ $\gamma = 0.683$, for the Manna model
in  $2d$  ($1d$ data did not show evidence of such corrections).
$f^*$ (see the caption) fluctuates about a constant 
over the optimum fitting interval, while the
effective $\tau$ varies strongly in the same.
The figure also has a 
pure power-law fit, with the estimate $\tau_s = 1.25$
\cite{biham}.  The latter yields a strongly curved function $f^* (x)$,
and shows that a power-law is not a good description.

The next candidate, to establish the validity of 
Eq.~(\ref{taufromdepinning}), is then the LIM model.
It also allows to use the geometry of the ``pinning
paths'', if the SOC and normal critical states are
directly connected. In directed percolation
depinning (DPD), 
the geometry of the random quenched landscape leads
to a self-affine picture of the interface dynamics,
described by the quenched KPZ equation
\cite{HHZ,Barabasi}. The interface motion
takes place so that the interface invades the voids
(constituting avalanches)
of multiconnected network of pinning paths, and
thus the sizes of the avalanches are related to the
void sizes. This is given by (in the LIM case) the
structure of an ``elastic percolation problem''
\cite{pinningpaths}, so that the the RHS of the qEW 
is always negative semi-definite,  $f \leq 0$. 
There are two fundamental issues: how the scaling of voids,
the relation of size vs. area is described with an appropriate
roughness exponent, $\chi_{loc}$, and what is the probability
to produce an avalanche if the interface is unpinned at a
particular spot (\cite{huber,zeitak,jost}).
In analogy with DPD \cite{huber}, it follows
that 
\begin{equation}
\tau_{s,dep} = 1 + (1/(1+\chi_{loc}))(1-1/\nu)
\end{equation}
which using the global $\chi \sim 1.20 \dots 1.25$, produces
\begin{equation}
\tau_{s,dep} \approx 1.08.
\end{equation}
This is the same as the prediction of Eq. (\ref{taufromdepinning}),
showing that assuming strict self-affine
avalanche scaling the SOC and depinning ensemble predictions
for $\tau_s$ (and other such exponents) coincide.

\begin{figure}[ht]
\centerline{
        \includegraphics[width=6cm]{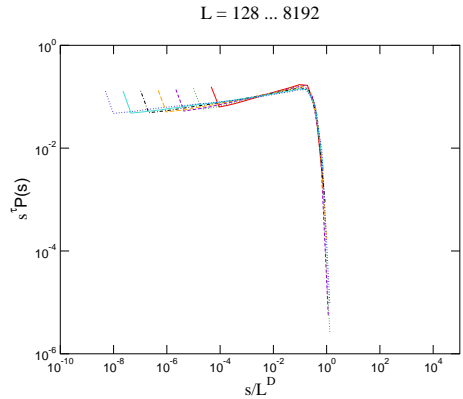}
        }
\caption{A data collapse using $\tau_s \sim 1.08$ and
$D_s \sim 2.20$ for the LIM-SOC model (Leschhorn-type
automaton, unpublished).}
\label{omafig}
\end{figure}

Figure \ref{omafig} shows a representative scaling plot
of SOC simulations of a $1d$ qEW/LIM, of the Leschhorn-type.
System sizes upto $L=8192$ have been studied, with 2 $\times$ 10$^6$ 
avalanches for the largest $L$. It is rather evident that
the bulk part of the distribution does not collapse that
well, for the largest system size the fitted 
effective $\tau$ is way off from the predicted 1.08
(1.024 for $L=8192$). An attempt to fit the exponents
to an asymptotic value and a correction implies
$\tau_{s,SOC} (\infty) \sim 1.115$, clearly off
the expected value.

Thus, in analogy to the  $2d$  Manna model, it seems that the
ensemble here is of importance, or in the interface language
the parabolic interface profile. It would be of interest
to attempt such comparisons in higher dimensions, where
one needs more exponents (or assumptions
to describe the probability distributions since the
avalanches have in addition to an area vs. volume
relation also a perimeter length vs. area
relation \cite{higherdim,zeitak}).
One would expect in any case that $D_s \leq d+\chi$ for the cut-off 
dimension, since the SOC state can of course not be 'over-critical'.

Surprises in trying to match the translation-invariant
non-equilibrium phase transition and the SOC one present
the question whether the differences vanish
asymptotically, so that the exponents of the statistically homogeneous 
ensemble are recovered? The other possibility is that 
the SOC ensemble {\em is an independent one}. Microscopically,
this would result from a non-uniform density of grains (or average 
force acting on the interface). In $1d$, a site $x$ will get a larger grain 
flux from its neighbor on the bulk side and a smaller one from the
boundary side - more trivially, there is a net flux of
grains towards the boundaries. 

This can be stated
a bit less casually by considering the integral of the deviation
from the force at depinning criticality, at $x$,
$\Delta F(x) = F_{dep} - F(x)_{SOC}$, where $F_{dep}$
and $F(x)_{SOC}$ are averages at the critical points
of the ensembles. This implies e.g. a finite-size correction to
the normal critical point,  $\int \Delta F(x) dx \equiv
\delta F(L)$. $\delta F$ of course follows from the exact scaling 
function of $F(x)_{SOC}$. In analogy to normal depinning,
this defines a correlation length exponent 
$\nu_{SOC}$ via $\delta F(L) \sim L^{1/\nu_{soc}}$.
Usual critical exponents like $\nu$ are derived from the RG,
assuming an ensemble with statistical translational invariance.
In the SOC case this is lacking, and it is not immediately
clear whether one can obtain the properties of the SOC state
from e.g. boundary criticality \cite{kent}.
In this respect the usual SOC models are more complicated than e.g.
the boundary driven Oslo model.

\begin{figure}[th]
\centerline{\includegraphics[width=0.7\linewidth]{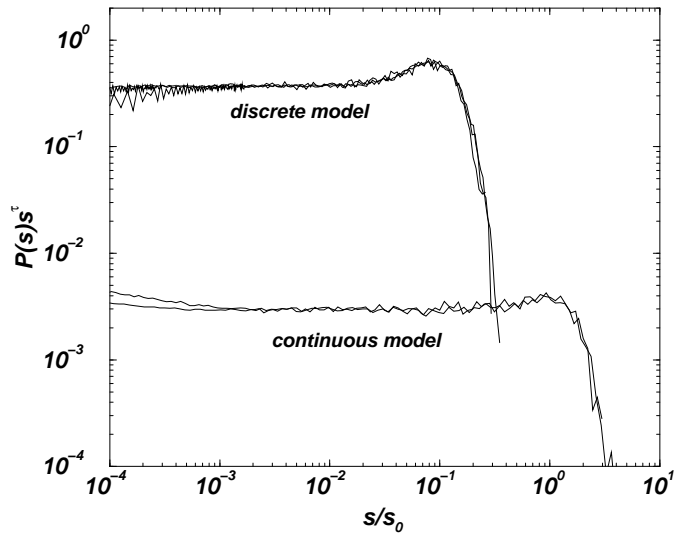}}
\caption{Data collapses
for system sizes between $L=128$ and $L=512$, for the
continous and discrete boundary-driven LIMs.  
The same value of $\tau_s=1.55$ suffices for both
\cite{pruessner}}  \label{Pruessner}
\end{figure}

Fig. \ref{Pruessner} shows, in contrast, a counter-example, from
the Oslo (ie. a boundary-driven LIM) model. Here the
continuum equation (of course integrated numerically) obtained
from a mapping of the sandpile to the LIM
is compared to the usual automaton for the avalanche size.
The result is that the exponents coincide, with each
other and with the analogous prediction for boundary-driven avalanches,
to Eq.~(\ref{taufromdepinning}) (that uses $\langle s\rangle \sim L$).
This examplifies the fact that the SOC state {\em can}
be understood, if translational invariance is present.

Finally, the properties of the SOC state are also visible
in the effective noise terms of the interface (e.g. $\tau$). 
It is not to be expected a priori, that the 
noise correlator and strength are translation:
consider the $\sigma$-noise, which reflects the tendency of the interface 
to move faster. This effect is the strongest in the center of a SOC
model, while with periodic boundaries the strength of the 
noise is independent of $x$. 
The origin of $\sigma$  depends on the exact sandpile rules,
in the Manna-model a site with $z_x=1$ can also get two grains 
from the same neighbor, making it overactive (note that one
can thus simply combine $\sigma$ and $\tau$ to a single 
quenched noise term), while in the Zhang, OFC and BTW models the
$\sigma$-field is set from the onset of an avalanche, including
the toppling order chosen. The probability for $\sigma(x,H) \neq 0$ 
is related to the {\it fluctuations\/} in the
density of active (and critical, $z_x = z_c$) sites
and is thus not accessible by graph theoretical \cite{Pri96}
or mean-field analysis \cite{vespignani-zapperi:1997}
of non-active configurations. It is
weaker in $d>2$ due to the ramified avalanche structure (and 
has no role with respect to the upper critical
dimension). Its strength evidently changes off the critical
point (since more of the neighbors are likely to be active
at the same time). 

\begin{figure}[tb]
%
\centerline{\includegraphics[width=6cm]{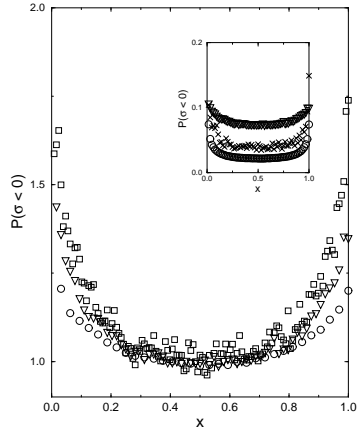}}
\vspace*{5mm}
\caption{$P(\sigma(x) <0)$ vs.\ $x$ scaled with $P(x=L/2)$ for various $L$ in
	the BTW model ($L=32$ (circles), $64$ (triangles), $128$
	(squares). Inset: for $L=64$ the same for BTW, Manna, 
        ricepile (triangles, circles, crosses) \cite{Alava2001}.}
\label{fig:P-sigma}
\end{figure}

Simulations of $\sigma$-fields in  $2d$  reveal that
the average noise strength depends on $L$, in terms of
$P_{L}(\sigma < 0)$, the probability to have
a non-zero value at a site at a toppling, and
$\left< \sigma(x) \right>$ is non-uniform in $x$.
The asymptotic BTW values are $\left<P\right> \simeq 0.081$ for all
and $\left<P\right>_d \simeq 0.121$ for dissipating avalanches.
Thus the dissipating avalanches have typically
stronger $\sigma$-noise. In the BTW model the
noise field has exponentially decaying correlations for
any fixed $x$, but with a decay length
that is largest in the center of the pile, and increases
in general with $L$. This and the existence of 
periodic oscillations in the correlations in the
bulk underlines the missing translational invariance
cite{drossel,ktitarev,barrat}.
It is unclear how the correlations of $\sigma$ relate
to the multiscaling in the model \cite{Teb99}, but
certainly they reflect also the columnar noise of the
interface equation \cite{columnar}. In contrast, for the Manna or LIM
automata the correlations in $\sigma$ decay rapidly.
Figure~\ref{fig:P-sigma} depicts a data collapse of 
$P(\sigma(x,y) \neq 0)$
along the cut $y=L/2$, $1\leq x \leq L/2$,  scaled with $P(x\simeq L/2)$,
evalued for those topplings that
are the last at the site during an avalanche. 
The probability increases at the boundaries with 
system size $L$. This shows that the lack of 
spatial translational invariance is a generic property
of the SOC ensemble.

\subsection{Off-criticality in SOC or interface models}

For interfaces the outcome of driving the
system ``too hard'', $F>F_c$ (or preparing a sandpile to be
``overcritical'', with $n_{tot}$), is just a cross-over
to, eventually, thermal noise on lengthscales $L \gg\xi $.
There are some typical
ways to push SOC models off the critical point. The first
option is to add a loss mechanism, to the dynamics of
grains, typically in the FES ensemble. The critical
state is then reached only for 
carefully tuned bulk dissipation
$\epsilon \sim L^{-2}$ (e.g. \cite{vespignani-zapperi:1997}).
The interface equation develops - using the same
projection trick as for the Manna etc. models - a linear
confinement term
\begin{equation}
        \frac{\partial H}{\partial t}\sim
\nabla^2 H + \eta(x,H) -\epsilon H(x) + F(x,t)
			\label{eq:H-eq-eps}
\end{equation}
where $\epsilon >0$ measures the strength of the dissipation.
The $-\epsilon H$ increases with a (small) probability only when
a site topples and the fluctuations in the dissipation contribute
to $\eta(x,H)$.
For e.g. the BTW model this means a perturbation, which is
irrelevant as long as the Larkin length associated with the cross-over
from columnar behavior is larger than $L$, and thus the avalanche
behavior is governed by the BTW dynamics. This is in fact the 
critical point of the constant velocity ensemble of depinning.
For all such sandpile
models one indication is however clear: for large avalanches the
critical behavior is cut off by the linear term, and thus 
any value of $\epsilon$ is sufficient in the TD-limit to bring
the system off the critical point (regardless of the ensemble)
\cite{dissipation}.
This is due to the avalanches becoming $d-1$-dimensional,
or that any site topples at most once. This result is of relevance
for the long-standing discussion about the existence of true
criticality with dissipation
in the OFC-class of models, but does not provide any hint
about any remaining scaling for $\alpha < 1/4$ \cite{ofcdissipation}.

\begin{figure}[bt]
\centerline{
        \includegraphics[width=6cm,angle=-90]{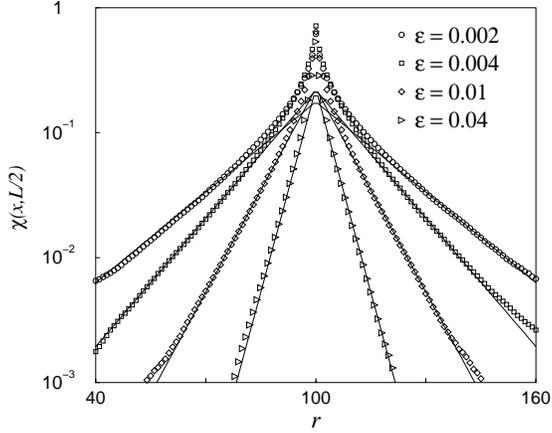}
        }
\caption{Time integrated response function  
         $\bar{\chi}_{h\to 0,\epsilon}$
         to a constant perturbation as a function of  $r$, the strength
of the perturbation.
         the linear lattice size is $L= 200$. The lines are exponential fits.
From ref. \cite{Barrat}}
\label{barfig}
\end{figure}

The driven sandpiles with boundary dissipation have been
also analyzed by various means. E.g. Kardar and Hwa attempted
to describe the grain flux dynamics by a RG treatment, to
understand the correlations in the ensuing steady-state
(for instance by its power-spectrum) \cite{hwakar}. 
The interface picture implies a cross-over between
quenched and thermal noise, which among others means
that the interface velocity can be related to the
fluctuations in the grain flux. Off the critical point
the avalanches overlap, and the stoppage of activity
becomes a rare event, like the interface velocity becoming
zero when $F>F_c$ \cite{oslooffcrit}. The expectation would be that
one obtains $1/f$-noise in the dynamics
\cite{zhang,krugnoise}. The correlations in the activity 
or  interface velocity reflect the Laplacian nature of grain
dynamics \cite{Barrat}. Figure \ref{barfig} shows an example of
the response functions of constantly driven BTW model, to
an extra perturbation. It exhibits clearly the properties
of a diffusive response to the extra drive field: the
response function ($\int^{t=\infty} \delta H dt$) decays
exponentially with $r$, distance from the location
where the perturbation was applied.

With the SOC criticality destroyed, the (random) drive 
and boundary dissipation still affect the fluctuations.
Usual measurements \cite{hwakar,Barrat,flu}) 
concentrate on instantaneous quantities like the local force/grain density
or activity/velocity. With a boundary condition $H=0$ imposed,
a drive $\langle F(x,t) \rangle = ft$ with $f$ a fixed
constant produces a constant average velocity which varies with
$x$, $\langle v(x) \rangle \equiv 
\langle \partial_t H(x,t)\rangle $. 
The continuum equation version is
\begin{eqnarray}
\frac{\partial H(x,t)}{\partial t} = \nu {\nabla}^2 H(x,t) + ft +
\delta f(x,t) + \eta(x,H(t)),
\label{eweqn}
\end{eqnarray}
a depinning ensemble with a
constant {\em drive rate} $f$.
This is not equivalent exactly to the normal
ensembles (constant force, or constant velocity \cite{Nara}).
The fluctuating part of $H$, $\delta H$, should reflect the fact
that the noise $\eta(x,v(x)+\delta H)$ develops temporal correlations 
that depend explicitely on $x$. Translational invariance is thus absent
also off the depinning critical point. Note that
the boundary regions are closer to depinning, 
and $\lim_{x\rightarrow \,0,\,L}v(x) = 0^+$,
while $\langle v \rangle  \propto f/L^2$.
In the proximity of the SOC critical point one feature
of Eq.~(\ref{eweqn}) seems to adhere to normal depinning:
the waiting times ($v(x)=0$) of the interface follow
roughly a power-law distribution $P(t_w) \sim t_w^{-1+\beta_w}$ \cite{pang},
where $\beta_w$ is the roughening exponent of the FES/depinning
problem, with no signs of logarithmic corrections \cite{dickmannotpl}
in the case of the Manna model \cite{laurson}.

\begin{figure}[f]
\centerline{\includegraphics[width=6cm]{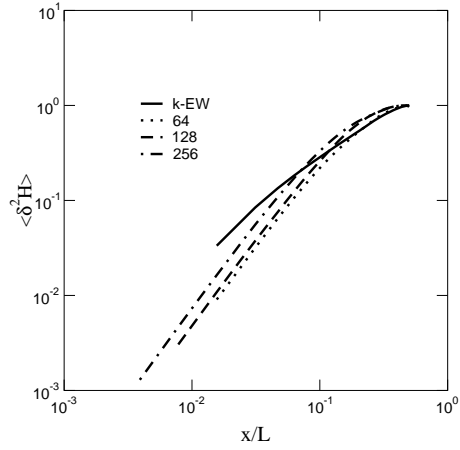}}
\caption{Amplitude of the interface fluctuations (normalized
 with $\delta H(x=L/2$) $\langle \delta^2 H \rangle (x)$ in $1d$ for both
the QEW and a thermal EW model with similar boundaries.
The drive rate, for grains, $f=1/100$ for the former,
for all the $L = 64 \dots 256$. \cite{AlaCha}}
\label{chatto}
\end{figure}

The $\delta H$ for a $1d$ interface is an example of return-to-zero 
stochastic processes recently analyzed by Baldassari et al. 
\cite{baldassari}. Here it is complicated by the structure
of $\eta(x,t)$. For the truly thermal EW equation, without
noise correlations the average fluctuations form a parabola
(in amplitude), which is easy to understand since the EW 
interface profile is a random walk with the given boundary
conditions. Measurements of the amplitude $\langle \delta^2 H \rangle_x$
as a function of $x$ for a $1d$ LIM SOC model, driven off the 
critical point, imply that the function changes from the EW form. Likewise,
the two-point temporal correlation function of the interface
has effective scaling behavior that combines the effect of
``kicks'' (drive by grains) and the relaxation of the interface,
and also reflects at the boundaries the fluctuations in the
grain flux, from the interior of the system.
All such details are awaiting real analytical understanding.
The fluctuation profiles in Fig. \ref{chatto}
show the same scaling behavior for all the system sizes
as a function of $x$ for the $1d$ QEW: for $x$ small
$\langle \delta^2 H \rangle (x) \sim x^{1.8}$, instead
of $\sim x^1$ as for a normal EW. 

\section{SOC in the Kardar-Parisi-Zhang depinning}
The second fundamental equation for depinning problems
is the quenched Kardar-Parisi-Zhang one \cite{DPD,toinen}, 
\begin{equation}
\label{quenched_kpz}
\partial_t H({\bf x},t) = \nu \nabla^2 H({\bf x},t)
+ \lambda (\nabla H({\bf x},t))^2 + F + \eta(x,H(x,t)),
\end{equation}
$\lambda$ measures the 
strength of the celebrated KPZ nonlinearity, proportional to
the squared local interface slope \cite{Kardar_1986}. The previous analysis
of reaching a non-equilibrium critical steady-state can now
be repeated in the presence of the $\lambda$-term, to study
the difference between the QKPZ and the EW equations 
(a non-linear Langevin equation and a linear diffusion equation) \cite{szabo}.
The dynamics of the latter is insensitive 
to initial profiles (unless no other noise is present),
while the QKPZ has a growth component $\partial_t H$
perpendicular to the interface slope, always non-zero. 

The Eq.~(\ref{quenched_kpz}) suffers from the problem
that it is notoriously hard to integrate numerically.
Some discretizations exist, a reasonable example \cite{numer} is given by
\begin{eqnarray}
f_i(t) = \nu \nabla^2 H_i (t) +
{\lambda \over 4}
\left( [H_{i}(t)-H_{i-1}(t)]^2 + [H_{i+1}(t)-H_{i}(t)]^2 \right)
\nonumber \\
 + F(t) + \eta_{i, H_i}.
\label{KPZdiscr}
\end{eqnarray}
where $f_i(t)$ is again a local force.
The {\em QKPZ sandpile} is such that a site gives $2\nu$
particles or units of energy to its neighbors once it topples, and the
toppling criterion also compares the integrated activity at site $i$
to its neighbors. In the interface picture
$H_i(t+1)=H_i(t)+\Theta(f_i(t))$. With this definition, $H(x,t) =
\int^t \rho(x,\tau) d\tau$ maps the discrete QKPZ equation
to the 'sandpile' and vice versa.
Now the simple dynamics $\partial_t H = \theta(f)$ 
and pinning boundary conditions $H=0$ at the boundaries imply a SOC ensemble,
if $F(t)$ is ramped slowly as in usual sandpiles. Note that
the exponents of the depinning transition depend on e.g.
if $\lambda$ approaches zero with $\Delta F \equiv F - F_c \rightarrow 0$ 
(One can map a directed sandpile to 
the KPZ equation, without considering the history
\cite{dirkpz}).

\begin{figure}
\centerline{\includegraphics[width=6cm]{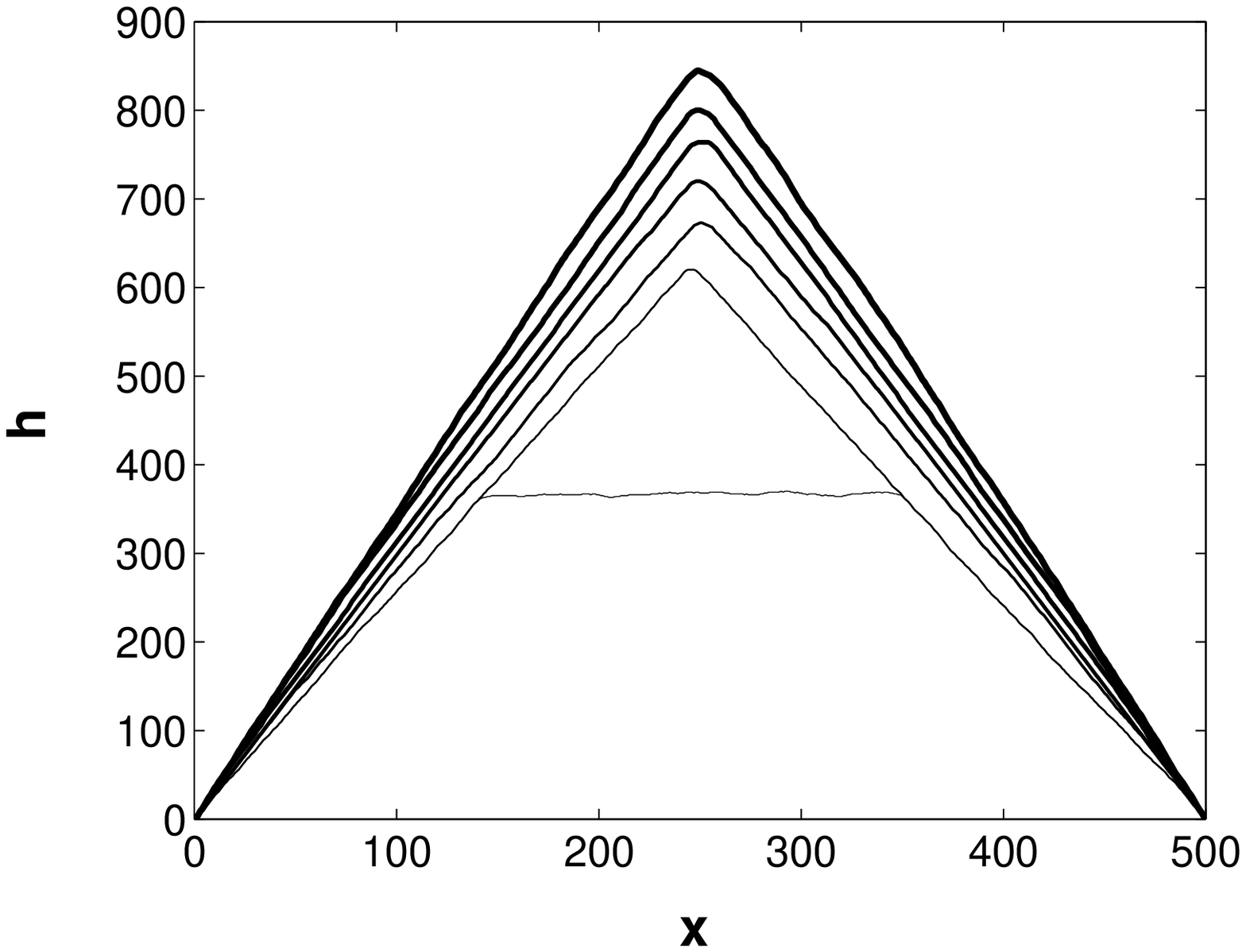}}
\caption{Typical interface configurations for a $1d$ QKPZ, with
SOC drive and boundary conditions, for a system of 500
sites. The one with the flat top is an unstable one from
the initial growth event before steady-state of the model,
and the others are from every 20000th time step
\cite{szabo}.
}
\label{snapshot}
 \end{figure}

The roughening, the 
development of average local slopes, depends on the sign of $\lambda$.
If $F(t)>0$ the interface becomes unstable if $\lambda>0$.
If $\lambda <0$ one can study a separation of time-scales:
the system is driven so that $F(t)$ is increased after an avalanche has
relaxed.
Fig.~\ref{snapshot} shows a series of snapshots from a simulation of
the QKPZ equation with $\eta_{i, H_i} =\pm g$ with equal
probability. This describes random toppling thresholds that change
after each toppling in the sandpile language. 

The shape of the interface is triangular in 1+1-D as expected
with $\lambda<0$ (see also \cite{santos}).
For a flat slope the surface tension is negligible, and a slope
arises from the balance of the nonlinearity and the driving force,
$\langle |\nabla h| \rangle = \sqrt{2 \zeta t/\lambda}$. 
This is analogous to the development of the
steady-state grain configuration in usual SOC.
In the ensuing phase usual avalanche properties can
be measured, eg. $\tau_s = 2.5 \pm 0.1$, implying
that the average avalanche size is independent of $L$, 
contrary to usual sandpiles with anomalous diffusion,
since $\tau_s >2$.

The avalanches are in addition time-translation invariant 
only on the average:
the largest avalanches happen when the roughness of the interface
is the largest. Local slopes, i.~e. $|H_i-H_{i-1}|$ 
alternate between $n$ and $n+1$ where $n$ is a positive integer. 
In spite of such deterministic trends there is no characteristic avalanche
size. Such dynamics (for the size, duration, and support) of 
avalanches eventually changes due to the development of 
a critical slope $m_c$. This is a signature of different
effective dynamics of avalanches, above $m_c$, which
one reaches in $L^2$ avalanches.

The combination of slow drive and boundary pinning eventually
pick a  symmetry for the average interface profile. 
Write the interface field on one of the sides of the triangle (1d) as
$H(x,t) = m(t)x + \delta H(x,t)$. Inserting this into the QKPZ equation
we obtain for $\delta H$
\begin{equation}
{\partial \delta H({\mathbf x}, t) \over \partial t} = \nu \nabla^2
\delta H + {\lambda} m(t) \nabla \delta H + {\lambda \over 2} (\nabla
\delta H)^2 + \eta({\mathbf x}, H),
\label{mKPZ}
\end{equation}
subtracting the mean-field solution, 
valid for intermediate times $\delta t \ll t$.
The effective interface equation has a linear term in
$\nabla \delta H$, that will dominate over the KPZ-nonlinearity. 
This is a full analogy of depinning in the presence of anisotropic
quenched noise \cite{TKD}, explaining the presence
of this regime.

\section{Conclusions and discussion}
In this article we have outlined the connection
of sandpile-like cellular automata, exhibiting ``self-organized
criticality'', to other non-equilibrium phase transitions.
Such mappings allow to establish a number of facts
about sandpiles, and leave a number of open questions.

The first observations concern the role of ensembles
and rules. The mappings to interface depinning reveal,
how SOC avalanche properties depend on the details of
the SOC automaton. The connections to both absorbing
state phase transitions and the depinning illuminate
the role of history effects in sandpile dynamics.
Below, a few evident conclusions are listed.

\begin{enumerate}
\item the noise in the LIM-representation settles the
universality class, in the absence of strong additional
symmetries (BTW model).
\item
the upper critical dimension, in the presence of a surface
tension, is four.
\item
for non-SOC ensembles the critical exponent follow from
the FES/depinning transition. In some cases these present
an open problem, since the QEW with certain kinds of
correlated noise is not well understood (in terms of
$\chi$, $z$ etc.). This has not been resolved in
the NDCF-picture either, due to the lack of analytical progress.
\item
the derivation of the avalanche exponents for SOC
sandpiles is an open problem. In some cases (Zhang,
Manna/NDCF class) this can be directly traced to
the fact that the noise of the interface equation
becomes $\xv$-dependent. Thus translational invariance
is broken. The same holds for off-critical SOC sandpiles.
\item
the off-critical models can still show criticality (eg. of 
the thermal EW-class)
\item
temperature can be introduced easily \cite{temp}, via the
idea of adding ``thermal noise'' into the interface
equation \cite{chauve1,creep}, and interpreting it in the sandpile language.
\item
the rule of conservation laws is highlighted
(Laplacian operator, the FES class and its description
via an activity picture).
\item
the boundary conditions of the SOC ensemble may 
(QKPZ ``sandpile'') lead to different
dynamics, due to symmetry-breaking.
\end{enumerate}
In summary, the existence of ``SOC'' is by such
mappings revealed to be a result of combination
of ``non-linearity'' (underlying {\em ordinary non-equilibrium
phase transition}) together with a slow drive in an inhomogeneous system 
(net current of particles or energy). It is not
an ``attractive fixed point'', but arises from
fine-tuning the drive rate so that avalanches separate.

One outcome is naturally the prospect of observing
``true SOC'' by preparing an experiment with boundary
conditions that exactly correspond to what the theory
would imply. Driven interfaces in random media (domain
walls in random magnets, for instance) would seem to
be an obvious candidate. Note that the predictions for
open systems differ depending whether the system is
exactly in the SOC state or not. 
Concerning theoretical aspects, the mapping of dynamics to a history
description might be of advantage in other problems,
like coupled map-lattices (for the contact process,
e.g. a new exponent arises \cite{dm,hist1}) and reaction-diffusion
systems (eg. \cite{systems}).

For SOC-like cellular automata and their continuum
descriptions one may list a number of open topics,
including those related to the
list above. Understanding the nature of the SOC
ensemble; the origins of multiscaling therein; 
mapping depinning to absorbing state phase transitions
\cite{AM2002}
renormalization techniques for both depinning (in 
non-standard ensembles like the SOC, and with correlated
noise) and the conserved field theory of NDCF models.

Other topics are the continuum descriptions of
models with more complicated rules (curvature 
-dependent thresholds, like for sandpiles derived
out of driven MHD equations \cite{solarfes}, in both SOC and 
FES ensembles; the vortex flow model of Paczuski
and Bassler that may be in the randomly
boundary -driven columnar noise LIM-class \cite{bassler}, the
non-local rules employed already by Kadanoff et al.
\cite{kadanoff-etal:1989} and so forth. In the
interface context, the presence of long-range interactions 
substituting the Laplacian surface tension in Eq.~(\ref{qew_eq}) 
changes the scaling exponents \cite{Tanguy_1999}. For all
such cellular automata one expects that they can be
mapped to known continuum equations with quenched
randomness, possibly with noise correlations. This
also implies that e.g the {\em predictability} \cite{pred} of
such SOC systems is very low and that the damage tolerance
is high \cite{staple}. An analogy is provided
by a percolation problem, in which one removes bonds around $p_c$ 
randomly till the spanning cluster is broken; then new ones
are added again till a spanning cluster is re-formed.
Clearly it is to first order difficult to ``predict'' whether
the largest cluster is system-spanning or not and in fact
the only information hidden in the behavior is the usual
spanning probablity (vs. $p$).

{\center \bf \Large Acknowledgments}
\vspace{0.5cm}

I would like to acknowledge in particular Kent Lauritsen for
collaborating on the connections between sandpiles and
depinning, as well as all my other collaborators on such
issues, including Miguel Mu\'noz, 
Romualdo Pastor-Satorras,  Ron Dickman, Alessandro Vespignani,
Amit Chattopadhyay, and Lasse Laurson. Stefano Zapperi is
thanked for a pleasant stay in Rome, where this manuscript
was being prepared, and the editors Elka and Rodolfo, for 
the possibility of contributing to the book. Also, Hugues Chat\'e
is thanked for the data of Table 2.
Financial support from SMC, La Sapienza, Rome, and
the Center of Excellence program, Academy of Finland
is also acknowledged.

\end{document}